# Data-driven strategic sensor placement for detecting disinfection by-products in water distribution networks

Aristotelis Magklis[1] and Andreas Kamilaris[1,2]

[1] CYENS Center of Excellence, Nicosia, Cyprus.
[2] University of Twente, Enschede, The Netherlands.

**Abstract.** Disinfection byproducts are contaminants that can cause long-term effects on human health, occurring in chlorinated drinking water when the disinfectant interacts with natural organic matter. Their formation is affected by many environmental parameters, making it difficult to monitor and detect disinfection byproducts before they reach households. Due to the large variety of disinfection byproduct compounds that can be formed in water distribution networks, plus the constrained number of sensors that can be deployed throughout a system to monitor these contaminants, it is of outmost importance to place sensory equipment efficiently and optimally. In this paper, we present DBPFinder, a simulation software that assists in the strategic sensor placement for detecting disinfection byproducts, tested at a real-world water distribution network in Coimbra, Portugal. This simulator addresses multiple performance objectives at once in order to provide optimal solution placement recommendations to water utility operators based on their needs. A number of different experiments performed indicate its correctness, relevance, efficiency and scalability.

## 1    Introduction

One of the most important needs of humanity is the access to clean drinking water. Water can contain various contaminants that may cause different health problems to the consumers. Providing civilians with purified water is a critical objective at every society, to ensure that consumers are not exposed to health risks[1]. To achieve such a goal, it is vital to understand which contaminants can be harmful for the human health and how to detect them before they reach the households. In this context, it is of outmost importance to detect pollutants and contaminants as early as possible, to stop any polluted water from reaching the public.

Disinfection byproducts (DBPs) are chemical compounds formed when disinfectants such as chlorine react with natural organic matter (NOM) in the water[2]. Some examples of natural organic matter include humic substances and fulvic acids that can alter the colour, taste or the odour of drinking water[3]. Apart from NOM, a number of environmental parameters affect the formation and proliferation of DBPs, such as temperature, turbidity, potential of Hydrogen (pH), topology and, in certain conditions, even climate change which leads to higher ambient temperatures, assisting in the formation of chlorinated disinfection byproducts[2].

In most cases, operators of water distribution systems (WDS) perform water sampling at the source (i.e., water depots) in certain frequency through the year (e.g., monthly), to adhere to local regulations of water safety. However, it is likely that DBPs can be formed later during the distribution of the water through the WDS, while interacting with NOM under certain environmental conditions, which may be different than the ones in the water depots. For this reason, the operators of WDS need to deploy sensory equipment throughout the distribution network, in order to capture the formation of DBP compounds as early as possible. To achieve that, operators can either employ expensive sensory equipment such as analyzers[4], or use less expensive methods, such as water quality sensors, that monitor the environmental parameters that form DBPs[5]. Thus, there are different types of sensors, that differ in cost and have different impact on the monitoring process. Lower cost sensors monitor water quality parameters, such as pH, temperature, total organic carbon (TOC), dissolved oxygen etc.[5], while real-time predictive analytics derived from those sensory measurements are essential for detecting anomalies or sudden changes in the parameter values that could indicate increased probability of formation of DBPs. On the other hand, high-cost analyzers can detect DBPs directly[4] allowing for more accurate detection, but their use is limited due to the high cost.

In this study, a software simulator, titled DBPFinder, is presented, which focuses on the strategic sensor placement for the detection of DBPs, supporting multiple performance objectives and simulated scenarios. DBPFinder offers various features, such as supporting input of custom environmental data and water distribution network topologies, simulation scenarios for multiple performance objectives such as minimization of time of detection and mass consumption, placement of sensory equipment based on a selected number of sensors, as well as randomized injections of DBPs for analysis of possible realistic contamination scenarios. DBPFinder allows to switch between objectives based on the needs of each water operator, for example to minimize the time of detection of the contaminants, or to minimize the amount of polluted water consumed by the civilians. Since each DBP family and compounds have different effects on human health[6], prioritization of sensor placement targeting certain DBPs plays a crucial role. Our work has been influenced by Kalita et al[7]., who attempted a prioritization of DBPs based on the factors mentioned above.

Our overall methodology constitutes a novel approach to the problem of strategic sensor placement for this particular type of contaminants, considering the constraints and challenges of the problem such as the limited knowledge about the impact of chlorine-based DBPs on humans and the precise conditions that favour their formation. Finaly, an important contribution is that we have published this simulator as open-source software[8], allowing operators of WDS around the world to use and adapt it to their precise needs.

## 2    Related work

Different studies have been conducted to understand the formation of disinfection byproducts, their impact on human health, trying to explain which environmental parameters and to which extent, affect the formation of disinfection byproducts. For example, it has been researched that chloroform (a Trihalomethane compound) was found carcinogenic in animal studies, which led regulators to limit THM levels in drinking water[9]. Epidemiological evidence has suggested weak but consistent associations between long-term exposure to chlorinated drinking water and higher risks of bladder, colon and rectal cancers[10]. Certain nitrogenous DBPs (e.g. haloacetonitriles) are known to be particularly toxic. Empirical studies report that haloacetonitriles (HANs), despite typically found in low concentrations, exhibit far higher cytotoxicity than regulated DBPs, such as THMs and Haloacetic Acids (HAAs).[11]

Most research related to disinfection byproducts focuses either on the treatment of these contaminants[12], or the predictive modeling[13] of detecting disinfection byproducts, not addressing yet the strategic sensor placement of equipment in a water distribution network towards their optimal detection. In typical WDN, a wide range of sensors are used to monitor a variety of water-quality parameters. Some of these parameters include residual chlorine[14] , pH[15], temperature[16], electrical conductivity[17] and turbidity[18]. These sensors range from simple low-cost ones[19] (pH, conductivity, turbidity) to high-cost analyzers[4] that can detect the exact concentration of certain disinfection byproducts. In practice, water utility operators use lower-cost sensors to monitor environmental parameters due to the fact that high-end DBP analyzers are expensive, bulky and inconvenient to move around, plus they require frequent laboratory maintenance and calibration. The assumption of most WDN operators is that after compiling enough data through the low-cost sensors, it is possible to develop predictive models to find the concentration of disinfection byproducts in the network with high confidence and acceptable precision.

Regarding strategic sensor placement[20], most literature focuses on other contaminants detected in the water, such as Escherichia coli[21] and organophosphates [22], or even on detecting leakages[23] based on the flow of the water in the network. In terms of algorithmic approaches for optimal sensor placement, a wide variety of algorithms has been proposed, such as genetic algorithms[24], the Non-dominated Sorting Genetic Algorithm (NSGA-II) algorithm[25], mixed integer programming techniques[26] and others.

Very few studies have focused specifically on DBPs. A handful of efforts have acknowledged DBPs as emerging contaminants for which strategic sensor placement is justified. An important study is the EPA's SPOT optimization tool, which considers water-quality issues [27]. Another study focused on the optimization of monitoring points for DBPs using cluster analysis, to provide monitoring solutions for detecting trihalomethanes and haloacetic acids [28]. The study from Ardila et al. [29] on the other hand, defines zones for each node based on hydraulic connectivity and considers two criteria: global representativeness and potential health risk [29]. This study also focuses on trihalomethanes and haloacetic acids.

A summary of the key algorithms applied for different contaminants or disinfectants is presented in Table 1 below.

*Table 1. Studies focusing on detecting contaminants and disinfectants and their corresponding algorithms.*

| Research Work | Algorithm utilized | Contaminant/Disinfectant | Metric(s) used |
|---|---|---|---|
| Weickgenannt et al. [21] | Non-dominated Sorting Genetic Algorithm II | Escherichia coli | Minimization of time of detection, minimization of risk |
| Salem et al.[30] | Genetic Algorithm | Chlorine | Minimize prediction error |
| Łangowski et al. [31] | Non-dominated Sorting Genetic Algorithm | Chlorine | Balance the number of sensors against estimation accuracy of chlorine levels |
| Ohar et al.[22] | Genetic Algorithm | Organophosphate pesticides | Minimize the number of exposed consumers |

| Shahra et al.[32] | Evolutionary Algorithm | Chemical/Microbial contamination scenarios | Maximize coverage and minimize time of detection |

In contrast, the present work provides a DBP-specific placement framework via our DBPFinder simulator. This simulator explicitly models the formation and transport of multiple DBP families in the network and optimizes sensor locations accordingly. Moreover, DBPFinder is released as open-source software, enabling water utilities to incorporate their own data and focus on their desired performance objectives.

## 3   Methodology

The methodology utilized for the development of our strategic sensor placement simulation software, is depicted in Figure 1.

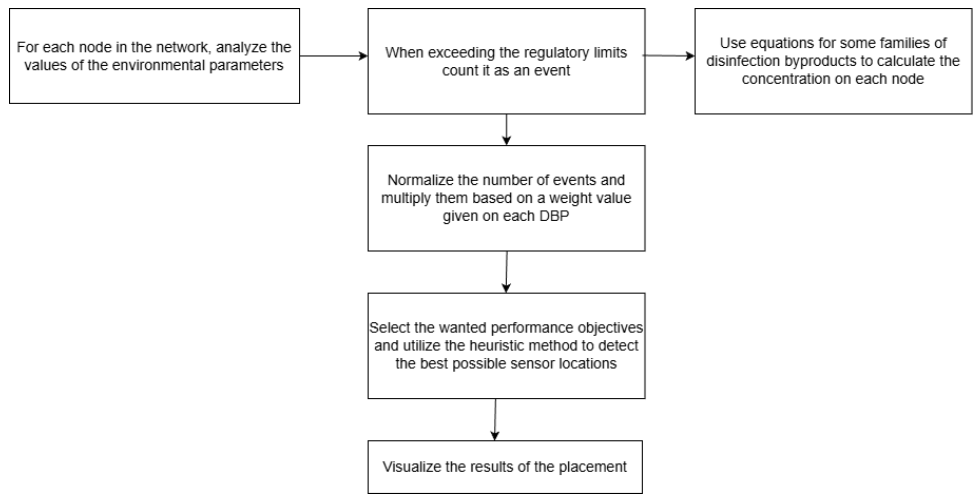

*Figure 1. Methodology*

The simulator, foremost, receives as input a water distribution network schematic that has been designed by some water utility operator. The network should be carefully designed based on the specifications that it has been built (nodes, edges, correct pipe lengths). Ideally, the operator should provide important information about the environmental parameters and their range of values, i.e., chlorine dosage, chlorine diffusion model, NOM through the network, etc.). Environmental data should include a wide range of relevant environmental parameters, estimated or recorded values at each node of the network at different time stamps. The more information available the better for the simulator to run properly. In real-world scenarios, we expect that information will become available only from certain locations within the network, which can change from time to time, when sensors are being relocated for better coverage. If the data provided by the operator is not adequate, for example, not enough data from multiple nodes in the network, or frequent measurements, the simulator generates synthetic data for the environmental parameters within specified ranges, in order to fill the gaps and construct a complete, usable dataset. The range of the values that are generated are calculated by taking the minimum value and the maximum value of the provided environmental parameters, in order to replicate a realistic dataset.

If possible, the network operators should provide a file containing the contracts of each node in the network, meaning the number of people that are impacted in each location of the distribution system. Furthermore, the thresholds of each DBP family are predefined. For instance, trihalomethanes are regulated at 100 μg/L in Europe[33], while thresholds for other DBP families are based on regulatory limits from other regions. The equations that have been implemented are derived from empirical studies and their implementation is based on the environmental parameters observed during lab experiments, as shown in Table 2. These equations have been selected, among others, as they presented high predictive accuracies. Additionally, a default time interval has been set to regulate the frequency of measurements, such as generating parameter values hourly throughout a week. The following sub-sections provide an in-depth analysis of the methodology, outlining each step, from data generation to performance objectives and lastly, accurate sensor placement depending on different contamination scenarios. Various contamination scenarios are explored, to highlight the range of possible risks to a WDN and the multiple possible locations for the smart deployment of sensors along the system.

### 3.1 Environmental parameters and values

Depending on the completeness of the environmental data provided by the water utility operators, some pre-processing needs to take place. The measurements taken by the operators are usually sparse, leading to small samples of data only. In order to have a complete dataset, measurements for at least every hour need to be taken, including a wide range of relevant environmental parameters, to be able to consistently analyze the concentration of DBPs. Furthermore, most data is captured from only specific nodes in the network, leading to lack of data across the whole network.

To mitigate this problem, the simulator analyzes the dataset given by the water utility operators, the number of nodes monitored, and the amount of data collected and proceeds by generating synthetic data. The synthetic data generation aims to cover these gaps and establish a comprehensive and usable dataset of water quality parameters that help in the formation of DBPs. Regarding the range of parameter values, based on the limited data provided by utility operators, the range varies from the lowest recorded measurement to the highest. A crucial environmental parameter is chlorine. Chlorine concentrations are calculated after the simulation software runs a chlorine decay simulation of the network under study using the EPANET software, which models the chlorine decay throughout the network to determine the concentration at each node[34]. The bulk decay model with respect to chlorine is described by the following equation, where C is the chlorine concentration, Kb the overall decomposition rate coefficient and *n* the order of the reaction with respect to chlorine:

$$dC/dt = -Kb \times C^n$$

In the table presented below, a portion of data records from a real-world dataset is provided, which includes environmental parameters and indicative values, timestamps of measurements, node identifiers within the network and the number of consumers that consist of each node using contracts, as we will see in the next subsection. A node in a WDN, is a point where water is delivered, stored or distributed. Nodes can be represented by junctions, tanks and reservoirs. A link in water distribution network is a connection between two nodes, through which water flows. Links can be the pipes of a network, the pumps and the valves.

Table 2. Example of environmental data.

| Timestamp | Node | Contracts | Chlorine (mg/L) | Temperature | pH | TOC (mg/L) | DON (mg/L) | BR (mg/L) |
|---|---|---|---|---|---|---|---|---|
| 20-10-24 0:00 | 1_1000 | 0 | 1.41 | 19.03 | 8.3 | 0.14 | 4.26 | 3.62 |
| 20-10-24 0:00 | 1_1001 | 5 | 0.72 | 13.05 | 7.4 | 5.77 | 12.86 | 4.36 |
| 20-10-24 0:00 | 1_1002 | 0 | 0.32 | 14.25 | 6.7 | 9.57 | 10.53 | 2.82 |
| 20-10-24 0:00 | 1_1003 | 12.5 | 1.47 | 12.12 | 8.2 | 9.87 | 10.49 | 4.81 |
| 20-10-24 0:00 | 1_1004 | 0 | 0.29 | 20.95 | 6.7 | 12.13 | 10.57 | 3.08 |
| 20-10-24 0:00 | 1_1005 | 2.5 | 1.36 | 15.62 | 7 | 1.06 | 12.43 | 4.88 |
| 20-10-24 0:00 | 1_1006 | 0 | 0.49 | 12.85 | 8.3 | 5.22 | 6.38 | 4.9 |

| 20-10-24 0:00 | 1_1007 | 5 | 0.93 | 23.09 | 7.1 | 0.46 | 2.04 | 4.47 |
|---|---|---|---|---|---|---|---|---|
| 20-10-24 0:00 | 1_1009 | 0 | 0.86 | 23.91 | 7.9 | 8.74 | 8.65 | 4.61 |
| 20-10-24 0:00 | 1_1010 | 0 | 0.8 | 21.67 | 7.1 | 10.38 | 13.8 | 2.87 |

### 3.2 Regression models for the formation of disinfection byproducts

For optimal sensor placement, it is important to predict the probability of the formation of certain targeted DBP families in different nodes inside the WDS. Water samples must be collected from various nodes in the network. Afterwards, lab experiments are conducted to derive equations that can be accurately incorporated into data analyses, imported to the software to calculate the concentration of disinfection byproducts and predict the most optimal sensor placement.

Different functional groups of DBPs interact differently with natural organic matter, based on different values of environmental parameters. For this reason, multiple regression models have been proposed from multiple authors of scientific papers [35], each suggesting different combinations of environmental parameters and theta values. For reasons of prioritization, a weight system has been implemented for prioritizing detectability of each DBP family, based on the work of Kalita et al. [7]. This technique allows to priorize sensor placement based on the family of DBP with the highest weight value. In the table below, some example equations are showcased, which have been selected from review studies [34] and can be used in the simulation software to calculate the concentration of each family of DBP.

*Table 3. Equations extracted from empirical studies that predict formation of DBPs.*

| DBP | Example equation | R2 | Reference |
|---|---|---|---|
| THMs | $THM = 10^{-0.038} \times (Cl2)^{0.654} \times (pH)^{1.322} \times (time)^{0.174} \times (SUVA)^{0.712}$ | 0.88 | Uyak et al.[36] |
| HAA9 | $HAA9\ (\mu g/L) = -345 + 1.695(Temperature) + 93.1(pH) - 226(UVA254) + 4.95(Cl2) + 5.66(NO_2^- -N) + 16.6(DOC) + 0.325(NH_4^+ -N) - 0.0693(Temperature)^2 - 6.41(pH)^2 + 190821(UVA254)^2 - 1.73(NO_2^- -N)^2 - 3.77(DOC)^2 - 0.01663(NH_4^+ -N)^2$ | 0.811 | Okoji et al.[37] |

| HANs | T-HANs = $10^{-1.065}$ (Br)$^{0.346}$(DOC)$^{0.369}$ ×(Cl2/DOC)$^{0.520}$ (t)$^{0.238}$ (Temp)$^{0.373}$ × (R2 = 0.943, p<0.0005, n = 36) | 0.943 | Hong et al.[38] |
|---|---|---|---|
| THMs | 0.04121 (TOC)$^{1.098}$(Cl2)$^{0.152}$(Br-)$^{0.068}$(Temp)$^{0.609}$(pH)$^{1.601}$(time)$^{0.263}$ | - | Sohn et al.[39] |
| HAA9 | 30.0 (TOC)$^{0.997}$(Cl2)$^{0.278}$(Br-)$^{-0.138}$(Temp)$^{0.341}$(pH)$^{-0.799}$(time)$^{0.169}$ | - | Sohn et al.[39] |

### 3.3 Filling the gaps in environmental data

The generation of synthetic data aims to fill the gaps of environmental data missing on certain nodes throughout the network, aiming to create a complete dataset for a complete simulation. For each node, data is generated at specified measurement intervals (e.g. every hour) over a defined period (e.g. a week). Regarding chlorine, running the simulation of chlorine decay through EPANET for 72 hours can yield satisfactory results of chlorine concentration on each node. This way, the value of the chlorine being distributed in the synthetic data is realistic. For the other environmental parameters, based on the measurements provided by the water utility operators, ordinary kriging was selected as the spatial interpolation method to fill the gaps for the dataset. Kriging explicitly models the spatial autocorrelation inherent in DBP formation, driven by changes in precursor levels and hydraulic residence time across the network. An example of this method is implemented in Scenario B, in Section 6.2

The number of sensors can also vary, depending on the desired outcome. Furthermore, the simulator includes an option to adjust the initial chlorine injection point within the network. In a realistic scenario, the starting injection point originates from a tank or reservoir and expands throughout the network, depending also on the type of the water distribution system (grid, ring, radial or dead-end layout) as well as the size of the network (i.e. the number of nodes within the district metering area). This approach allows the randomization of the injection point, potentially selecting any node in the network. This can lead to multiple injection points at the beginning of the simulation.

### 3.4 Water distribution network

A calibrated water distribution network is required to be provided by water utility operators as input to the software in order to model and analyze the placement problem, as well as visualize the results of the placement. A calibrated network contains information gathered from measured values from the actual system. Thus, a WDN developed in EPANET contains pump and valve settings, demand patterns and pipe roughness coefficients that have been adjusted to match observed field data. Since access to real-world calibrated networks is limited, identifying DBP concentrations through simulations becomes even more of a challenging task, due to the lack of ground-truth network data (pipe sizes, water travel time, node placement). The calibration of the model of the network assists in the diffusion of the chlorine throughout the system, enhancing the accuracy of calculations for the concentration of DBP families at each node of the network. Lastly, the water utility operators can also provide a list of the consumers at each node in the network, allowing to solve the placement problem based on the performance objective related to the minimization of consumers' consumption.

### 3.5 Strategic sensor placement algorithm

By taking into consideration all data provided by water utility operators (i.e. WDS topology network, environmental parameters and values, regression models per DBP family – optional), we proceed with the simulations regarding the strategic placement. If the dataset is not sufficient, due to infrequent measurements and/or samples from only a few nodes of the WDS, before the strategic sensor placement, synthetic data is generated to fill the gaps of the dataset (Section 3.3). To consider a dataset as incomplete, the measurements of the water quality parameters must be scarce (e.g. every few months) and only from a very small number of nodes (e.g. 30% of total number of nodes). Afterwards, a chlorine diffusion simulation is executed, in order to determine the time of detection of chlorine, as well as to analyze the chlorine concentration at each node. Then, the simulator filters the nodes of the network and detects the best possible solutions based on the performance objective or objectives selected. Depending on their concentration, time of detection and contracts per node, the simulator analyses the most optimal locations based on the data provided. The final step includes the visualization of the placement of the sensors, based on the performance objectives specified by the user, whether employing a single-objective or a multi-objective approach, which are showcased in Section 4.7. In the case of a multi-objective approach, the model identifies the node most frequently selected across all performance objectives, with the results displayed using a pie chart. To solve the placement problem, integer programming is usually used, with certain heuristic options considered in very large WDS for faster execution. In certain cases, genetic algorithms (GAs) have been used, as seen in Section 2 and can also be applied to provide solution to optimal sensor placement. GAs iteratively evolve a population of candidate solutions through selection, crossover, and mutation operators.

# 4    The DBPFinder Simulator

Following the methodology described in Section 3, in this section we describe DBPFinder, an open-source web-based simulation tool, designed to perform strategic sensor placement in chlorinated water distribution networks, aiming to optimize the detection of disinfection byproducts (DBPs), formed when chlorinated water interacts with natural occurring matter. DBPFinder basically implements the methodology described in Section 3, employing integer programming as the algorithm to solve the optimal sensor placement based on the configurations/constraints/needs of each scenario. Developed with the Streamlit python library, using EPANET hydraulic models, DBPFinder integrates a wide number of environmental parameters and data, as well as customizable DBP predictive regression models/equations, in order to identify the most effective locations for sensor placement, towards monitoring DBP families like trihalomethanes (THMs) and haloacetic acids (HAAs). The simulator allows the users to input network models along with environmental data and to define or select DBP formation predictive models for tailored analysis. Furthermore, it is possible to include population data to minimize DBP exposure risk, as well as to consider multiple performance objectives, such as time of detection, normalized concentration score and event frequency for specific DBPs, for more targeted strategic placement. In the next subsections, we explain how the tool works, in relation to the above.

## 4.1    Data import

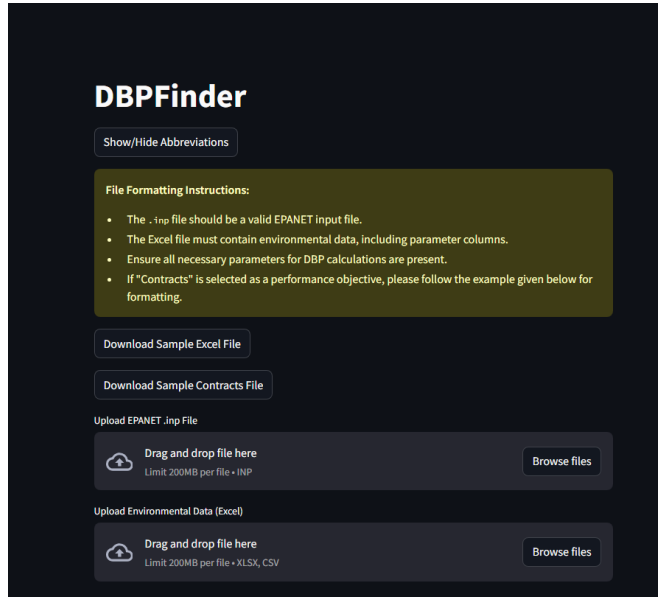

*Figure 2. DBPfinder data import menu.*

In order to run properly the strategic sensor placement simulation, the user must input a water distribution network that has been developed in EPANET, by clicking the 'Browse Files' button next to the 'Upload EPANET .inp file' option, as well as environmental data that corresponds to the template provided by the simulator, which organizes environmental parameters in columns and timestamps/values in rows. DBPFinder allows the user to download an example of an environmental data file, to guide the users on how to construct the input data accordingly. A similar template is used for the population affected per node of the WDN under study (Contracts). Figure 2 shows a screenshot of the DBPFinder tool, indicating where the data needs to be imported.

## 4.2 Disinfection byproducts formula

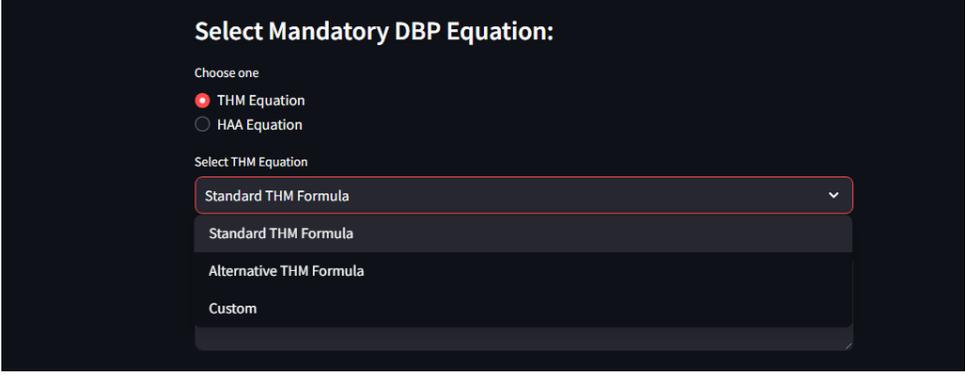

*Figure 3. Showcase of disinfection byproducts equations.*

Regarding predictive models used to calculate/estimate the concentrations of DBPs in different nodes of the water network, DBPFinder includes two recommended linear regression models, for trihalomethanes and for haloacetic acids, derived from Sohn et al.[39]. Furthermore, the tool allows the user to insert custom variables (i.e. additional environmental parameters) to create a custom formula, enabling the water utility operators to have more flexibility on the prediction of the concentrations of DBPs in their network, towards more optimal strategic sensor placement. This is visualized in Figure 3.

### 4.3 Performance objectives and prioritization

*Figure 4. Performance objective selection, weight system and sensor number selection.*

The performance objectives supported by the tool are the following:

● Minimization of time of detection, aiming to place sensors as close to the injection point to detect DBPs quickly.

● Minimization of mass consumption, aiming to reduce the consumption of polluted water by placing sensors in nodes that have lots of consumers, based on the contracts provided by the water utility operators.

● Placement based on concentration score, aiming to place sensors at nodes that have very high concentration of DBPs.

● Placement of sensors based on specified families of DBPs, aiming to focus on placing the sensors on nodes that exceed the regulatory limits of certain DBPs often.

Figure 4 shows the interface of the tool allowing to choose different objectives. The users are able to select multiple performance objectives at once, depending on the data they have available. For example, to use the objective related to minimizing the mass consumption, it is vital to include a contracts file. Two of the performance objectives are mandatory for the software to run, which are: (i) normalized score and (ii) time of detection. How these objectives are addressed analytically is presented in Section 5. The last two objectives include the detection of specific families of DBPs only, which can assist water utility operators who might be facing challenges with only certain DBPs. Moreover, the user can select which DBP should be prioritized when it comes to the strategic placement, by adjusting the weight of each family of DBPs. (see Figure 4). Weights are defined in the formula that calculates the concentration of each DBP and can be adjusted in the simulator to prioritize placement for specific disinfection byproducts when multiple ones are detected/predicted on some network at some particular point. This is further explained in Section 5.

### 4.4 Solving the optimization problem

The simulator works as follows:

In order to generate the final normalized score for each node, meaning which node should be prioritized for sensor placement, the steps are as follows. The number of disinfection byproduct events for each node are calculated:

$$total\_(THM\_(events\_node)) = node\_data\ [^{\prime} THM\ event^{\prime}].sum()$$

Based on how long the measurements are happening for, the event counts are normalized to the maximum possible number.

$$maxeventnodes = len(timestamps)$$

$$normalized_{thm_{event_{percent_{node}}}} = np.round\left(\left(\frac{total_{thm_{events_{node}}}}{\max{possible_{events_{node}}}}\right) * \frac{100}{2}\right)$$

After normalizing the event counts, the weight of each disinfection byproduct family is also multiplied based on the normalized event percentages.

$$weighted_{thm_{score_{node}}} = np.round\left(normalized_{thm_{event_{percent_{node}}}} * THM_{weight}, 2\right)$$

To finally achieve the total score node, the weighted scores for each disinfection byproduct are summarized and added to the dataset.

$$total_{node_{score}} = np.round\left(weighted_{THM_{score_{node}}} + weighted_{HAA_{score_{node}}}, 2\right)$$

After compiling all the necessary data, the strategic sensor placement process begins, employing integer programming. Nodes are filtered based on their node score. For instance, nodes with a score of 0.9 or higher, indicate a high concentration of disinfection byproducts. The values at each node are used to then address the performance objectives by filtering out a list with the candidate locations and selecting the appropriate nodes depending on the performance objective selected

### 4.5 Number of sensor devices

Lastly, it is possible to select the number of sensors to place in the network, as well as an option is given to randomize the injection point of the contaminants, for rapid prototyping of "what-if" scenarios.

### 4.6 Events for detecting specific families of DBPs

To support a wide variety of contamination scenarios and strategic placement possibilities, *events* are also a feature of DBPFinder software. An *event* happens when the concentration of the disinfection byproducts is above the regulatory limits and is described as shown below:

$$THM\_event = int(TTHM > THM\_threshold)$$

Harnessing the concept of events, the simulator can also adjust the placement for specific types of DBPs. By considering the number of events for each family of DBP per node, the placement can differ depending on the prioritized DBP. The flexibility of this approach is that it allows both multi-objective and single-objective approaches, accommodating computational constraints. This is particularly important for large-scale networks with thousands of nodes, where evaluating all the performance objectives at the same time may be time-consuming, thus achieving good spatial scalability.

### 4.7 Visualizations

The software allows the visualization of the placement of the sensors in the network, the name of the nodes the sensors have been placed and the corresponding performance objective result. You can observe two example visualizations in Figures 5 and 6 below.

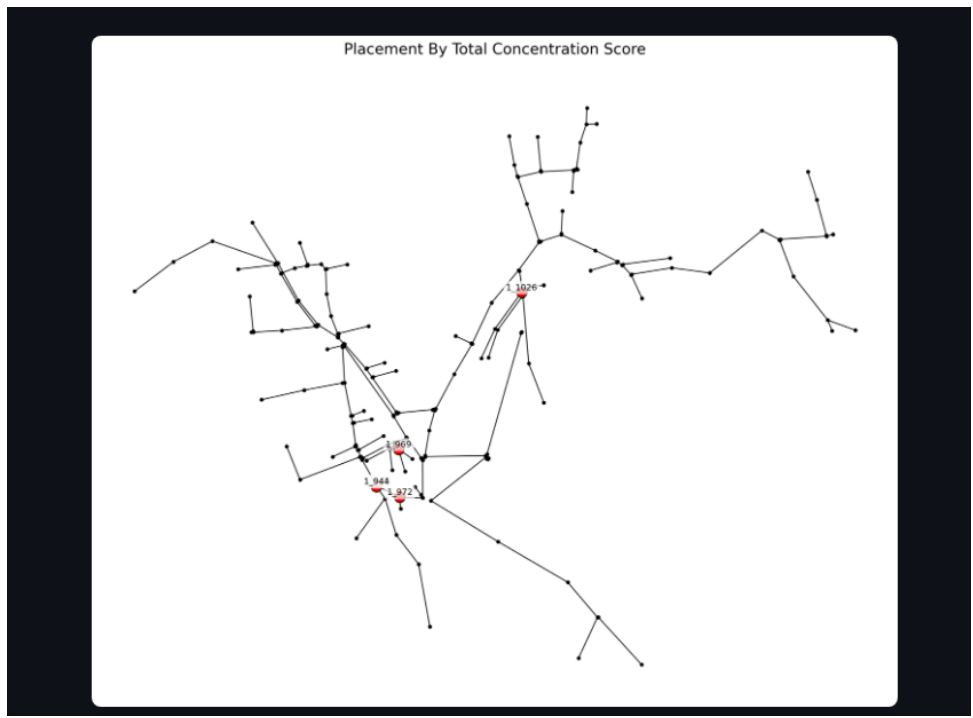

*Figure 5. Example of placement of sensors on a network through DBPFinder*

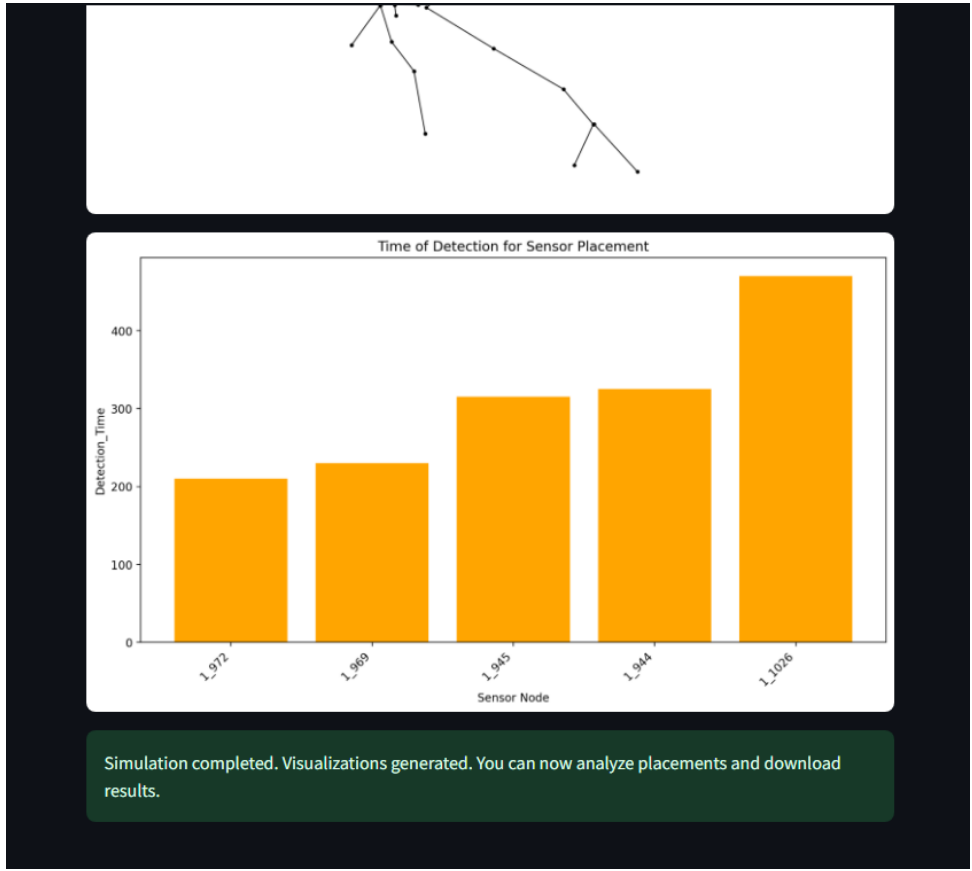

*6. Example of bar graph results from sensor placement through DBPFinder*

## 5    Case study: Coimbra, Portugal

The methodology was applied to a water distribution network at the civil perish Ameal e Arzila in Coimbra, Portugal (see Figure 7). The WDN provided by the utility operators is of moderate size, includes 227 nodes, one tank and one reservoir and is categorized as a dead-end network. A dead-end network, also known as a tree system, is characterized by a main pipeline running through the centre of an area with sub-mains branching off. These branch lines end without being connected to other branches, thus creating dead-ends.

The dataset provided by the operators of this WDN consisted of the following environmental parameters: chlorine dosage, pH, total organic carbon and temperature. By assigning a specific range to each parameter (e.g. chlorine range = 0.1, 2 mg/L), it is possible to generate distinct values for all the nodes in the network, for each measurement. In this specific case, synthetic measurements were conducted every hour for one week at all network nodes. To create a realistic approach for the values of the chlorine ranges, a chlorine decay simulation through time (72 hours) was run using EPANET [34]. This way, it is possible to analyze the chlorine concentration on each node in the network and detect possible candidate nodes that have high levels of chlorine. For the rest of the water quality parameters, the measurements of the utility operators were quarterly, meaning every 4 months and consisted of only 3 nodes, near the dead-end parts of the network. To fill the gaps regarding the rest of the parameters, the spatial interpolation method was utilized, which is explained in Section 3.3.

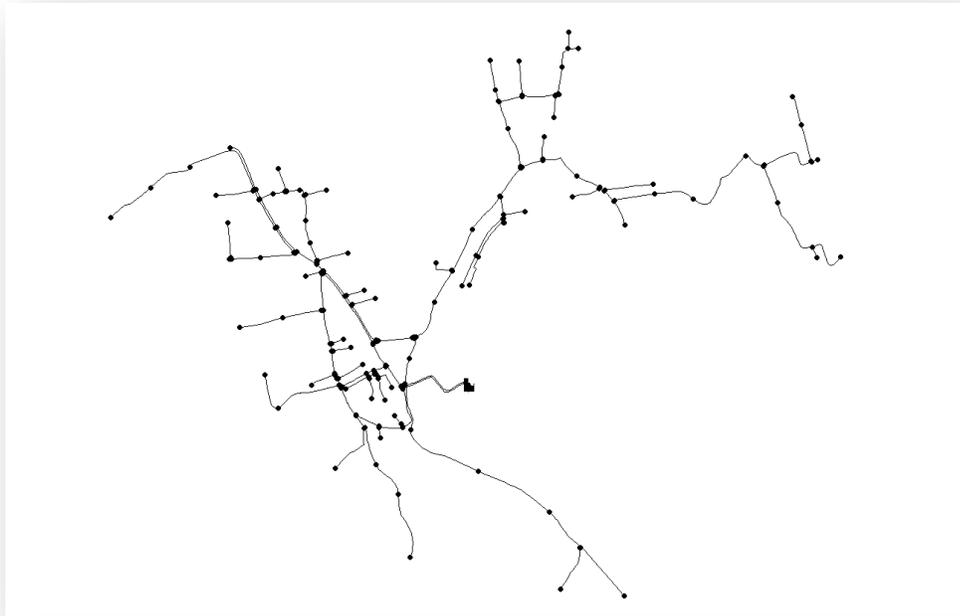

*Figure 7. Water distribution network.*

Furthermore, a weight was assigned for each DBP family that needs to be detected to prioritize the placement of the sensors, based on the precise needs of the particular water utility operators of Coimbra. Then, the equations employed for calculating DBP formation, derived from the work of Sohn et al.,[39], were used to predict the probability of formation those DBP families in the water network. The weights of the DBPs are listed in Table 4 and the equations utilized in this particular case study are listed in Table 5.

Table 4. Weight values for disinfection byproduct families.

| Trihalomethanes | Weight value: 0.4 |
|---|---|
| Haloacetic Acids | Weight value: 0.3 |

Table 5: List of predictive regression models applied in the case study of Coimbra.

| THMs | $0.04121 \, (TOC)^{1.098}(Cl2)^{0.152}(Br-)^{0.068}(Temp)^{0.609}(pH)^{1.601}(time)^{0.263}$ |
|---|---|
| HAA9 | $30.0 \, (TOC)^{0.997}(Cl2)^{0.278}(Br-)^{-0.138}(Temp)^{0.341}(pH)^{-0.799}(time)^{0.169}$ |

For the implementation of time of detection and mass consumption, two different functions were implemented. Foremost, a chlorine decay simulation ran through EPANET. As the chlorine reaches each node, the corresponding arrival time is recorded in the dataset, allowing the visualization of the time of detection. Regarding the mass consumption, the water utility operators provided another dataset that includes the number of contracts for each node, which serves as a proxy for the population affected by contaminants at each node (see Sections 3.1 and 4.1). The contracts are assigned to the nodes automatically at the start of the simulation, to include the performance objective of "minimization of mass consumption" to the list of outputs of the simulation. The users can also adjust the number of sensors to place in the network and the injection point, as discussed in Section 4.5.

The first performance objective tackled was the one that focuses on the nodes with the highest concentration, meaning the model filters out the remaining nodes with a value of 0.9 and above (see Section 4.4), and starts from the highest concentration and proceeds to the lowest, placing the sensors. This way it is possible to evaluate the best possible locations for sensor placement based on the concentration of DBPs per node. The next two performance objectives focus on minimization of time of detection and mass consumption.

For the time of detection, based on the chlorine decay simulation results, DBPFinder identifies the filtered nodes with the shortest detection times and places the sensors accordingly. Similarly, for the mass consumption, the simulator places sensors on nodes with the highest number of contracts, indicating nodes with the highest impact on human population (i.e. consumers exposed to the contaminants).

This way, this simulation covers contamination scenarios that include multiple DBPs (trihalomethanes and haloacetic acids), can also suggest solutions for specific families of DBPs based on their concentration levels and can also assist with the minimization of time of detection and mass consumption.

## 6   Evaluation

In this section, three different evaluation scenarios are considered for evaluating the DBPFinder simulator. The first two scenarios investigate the correctness and usability of the DBPFinder simulator, based on the case study of the Coimbra, Portugal. In the first scenario, the injection point of chlorine is at the tank of the network, using a small number of sensors. This is the realistic, real-world case of the WDN of Coimbra. In the second scenario, the chlorine injection point varies, and the environmental data is randomized, adding explicitly and purposely high values to some parameters to exceed the thresholds and trigger the formation of DBPs in different sections of the distribution network. Lastly, the third scenario focuses on the scalability of the simulator and how it behaves when requesting a very large number of sensors in a very large network. All three scenarios and their respective experiments are listed below, in the following subsections. It is assumed that the sensor devices deployed strategically in the network operate without any errors.

### 6.1   Scenario A: Realistic scenario

The water utility operators at Coimbra have calibrated the network to include an initial value for the spread of chlorine through the network [34]. Afterwards, quarterly reports (measurements of water quality parameters every three months) of the ranges of measurements of the environmental parameters listed in Section 3.1 and Table 2 were provided, for specific nodes of the network. For the rest nodes of the network, synthetic data was generated based on the method described in Section 3.3 for filling the gaps.

In the first experiment, we studied the relationship between number of sensors and "expected time of detection" for the respective performance metric. For this scenario, we assume that a minimum of five sensors are available to be placed throughout the network, based on the capacity of the actual Coimbra WDN. At the same time, we simulated a varying number of sensor devices, to understand the relationship and trade-off between performance of the metric and the number of sensors used, with the overall goal to detect contaminants as fast as possible. Figure 8 below showcases the results of this simulation. It is evident that 60-80 sensors have an important impact on the performance, while a larger number of sensors increases to a much lesser extend the overall performance. This also tells that with 60 sensor devices deployed in this network, expected detection time can be reduced to less than 5 hours.

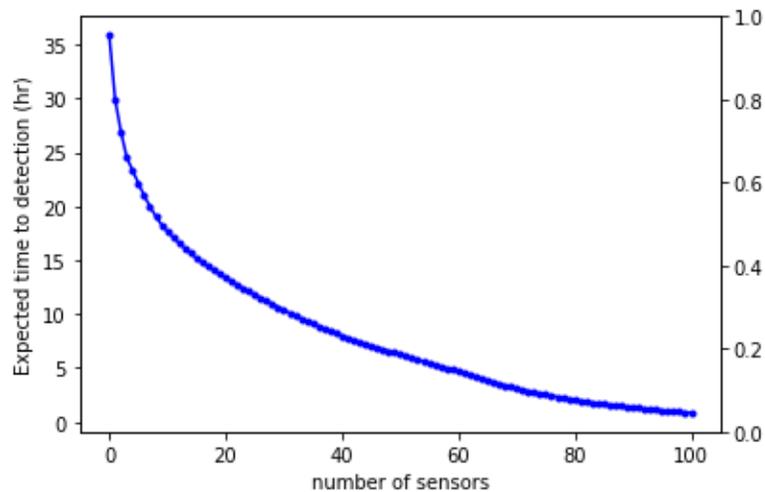

*Figure 8. Pareto Front for maximum coverage of the network and minimization of time of detection.*

Figure 9 visualizes the results of the strategic sensor placement for the specific case of 5 sensor devices. This decision was based on the normalized score of each node, as produced by the DBPFinder simulator, highlighted in Figure 10 (i.e. the nodes with the highest normalized score are shown). For this experiment, the expected time to detection based on 5 sensors is 35 hours, as depicted in Figure 8.

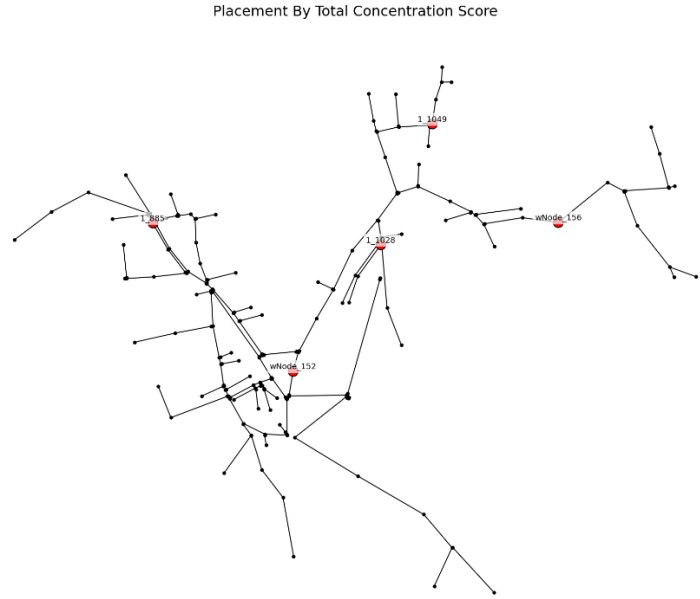

*Figure 9. Placement of five sensors based on normalized score*

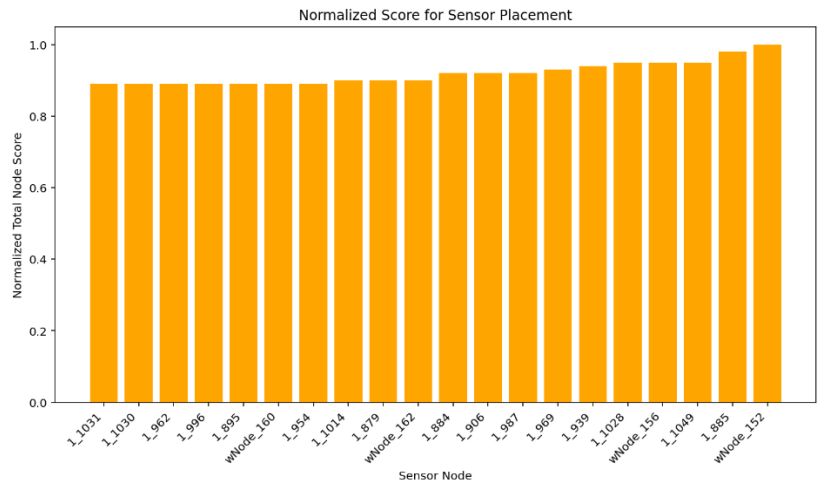

*Figure 10. Varying nodes based on their normalized score.*

In the second experiment, we examined the case of varying numbers of consumers exposed to contamination (i.e., placement by contracts), aiming to minimize the consumption of polluted water by civilians. Considering again the case of 5 sensor devices as the previous experiment, the suggested placement is presented in Figure 11. Even though the placement looks sparse, proposing placement to nodes that are close to the dead-end of the network, thus having lower concentration of DBPs, the simulation makes sense, because the number of contracts at those nodes is large, i.e. there are multiple consumers affected by those nodes. Some of these nodes have 25 contracts or more, leading to a high number of consumers, as showcased in the bar chart of Figure 12. All the nodes listed in Figure 12 are affected by the placement of the five sensors based on the suggested configuration.

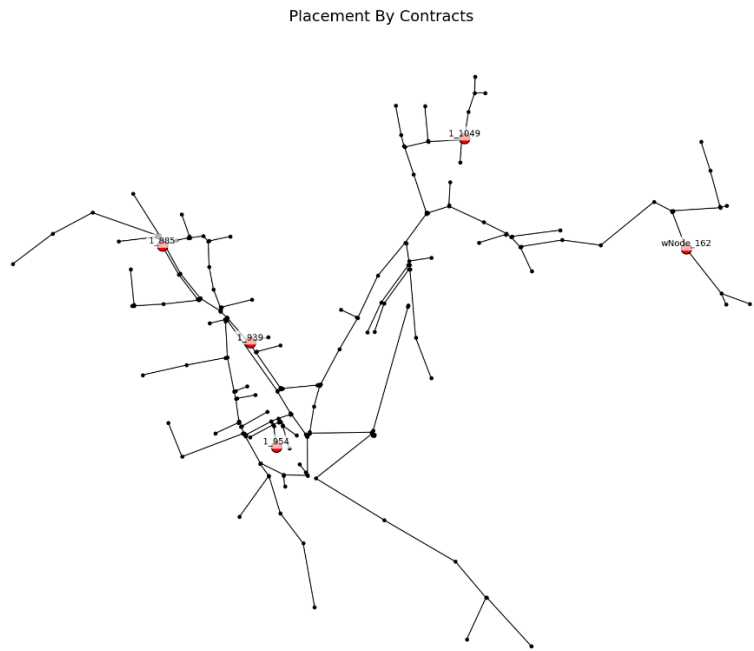

*Figure 11. Placement of five sensors based on contracts on network*

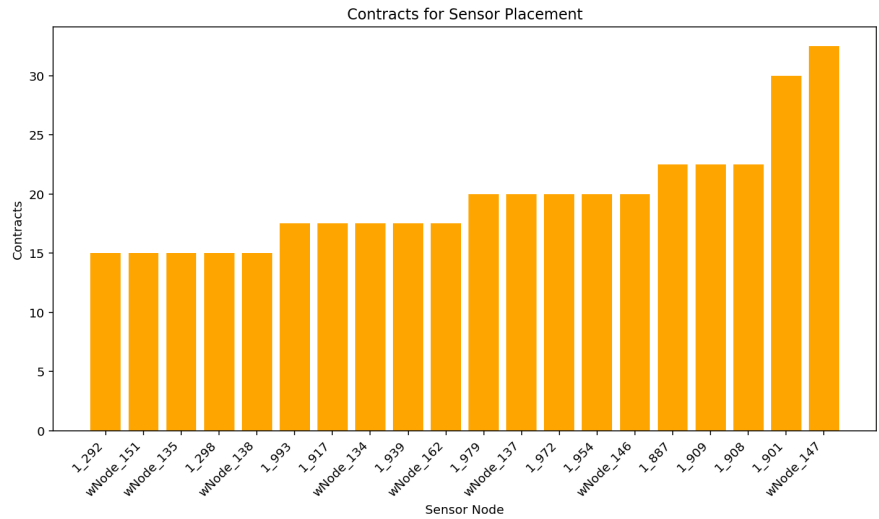

*Figure 12. Contracts per nodes in the network.*

The third experiment is about minimizing detection time, when only 5 sensors are available. In this case, the DBPFinder suggests that the sensors are placed near the injection point, at a small distance from one another, as shown in Figure 13. The detection time is calculated in minutes, and the nodes are selected from the filtered list of nodes that contain a total node score of a value of 0.9 and higher. The value 0.9, is translated into the lowest possible detection time, which is five minutes from the injection point. In the the case of the pilot study of Coimbra, injection point is the water tank.

Figure 14 shows the nodes affected by the placement, and it is a strong indication that DBFinder performed well, as the detection time at those nodes is less than five minutes.

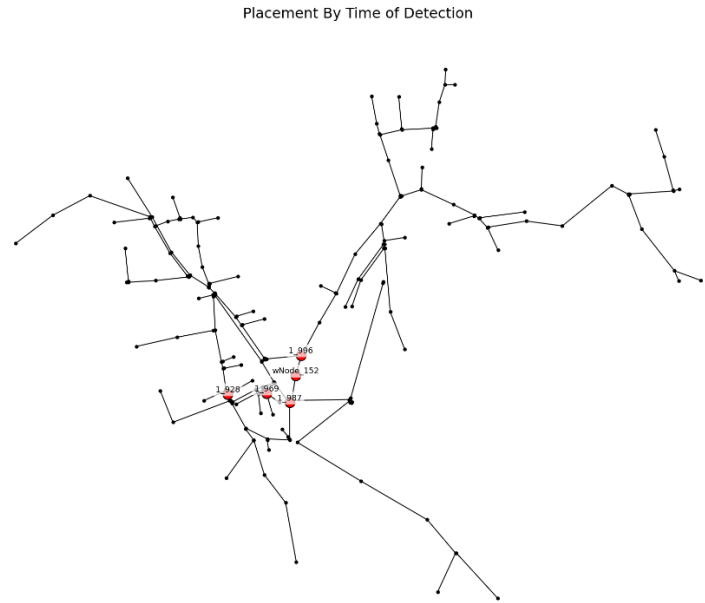

*Figure 13. Placement of five sensors to minimize the time of detection of DBPs*

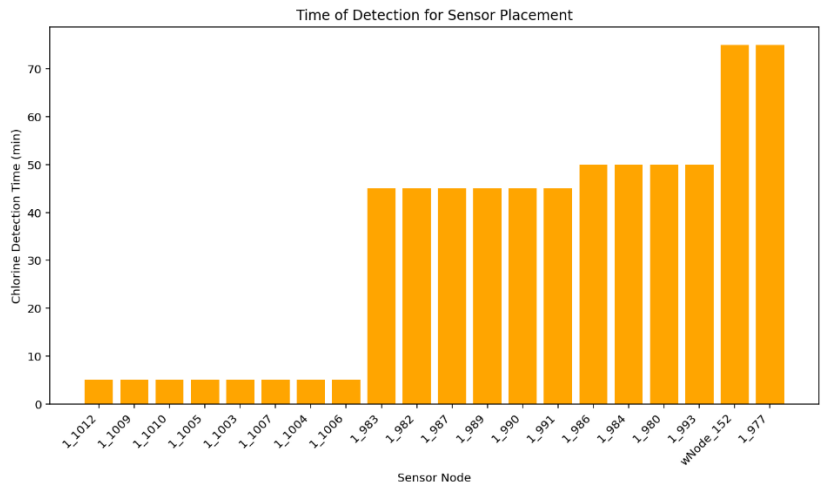

*Figure 14. Time of detection in minutes varying within nodes*

Furthermore, another experiment focuses on the prioritization of the detection of specific DBP families, as analyzed in Section 4.6. An event occurs when the concentration of some DBP family exceeds the threshold of the regulatory limits. This placement allows the water utility operators to focus on individual DBP families, allowing for more flexible placement scenarios, based on the specific monitoring needs of each water distribution system. In Figure 15 and Figure 16, the placement of five sensors for the prioritization of the detection of THM and HAA respectively in the network of Coimbra is showcased.

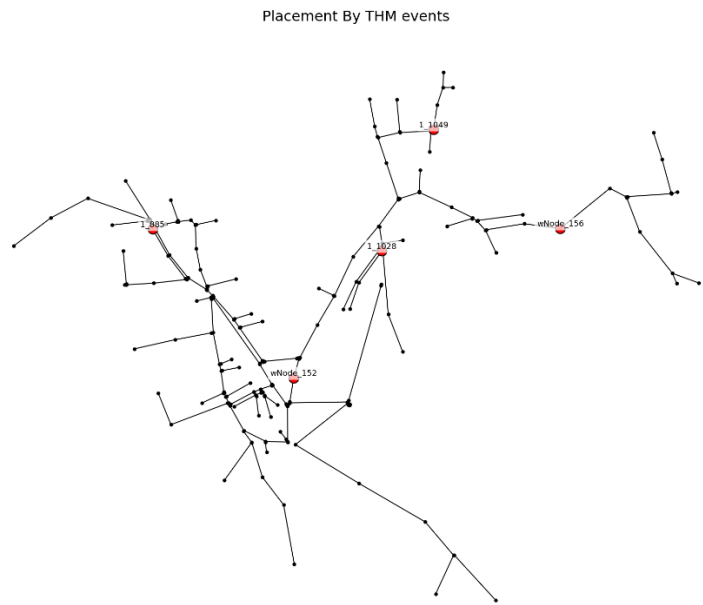

*Figure 15. Placement of five sensors based on exceeding thresholds for trihalomethanes*

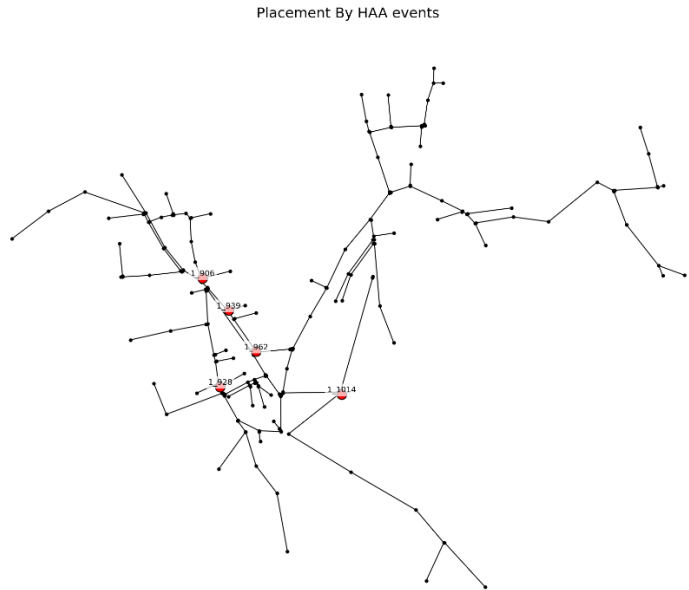

*Figure 16. Placement of five sensors based on exceeding thresholds for haloacetic acids*

Lastly, a summary of the most "popular" nodes in the network are highlighted, visualized via a pie chart. The nodes that are presented in this pie chart come from all the scenarios covered previously. This allows an in-depth analysis of the nodes that have been selected in all cases with different performance metrics considered. This allows the strategic sensor placement in certain nodes if the sensory equipment is quite limited and when many objectives need to be satisfied in parallel.

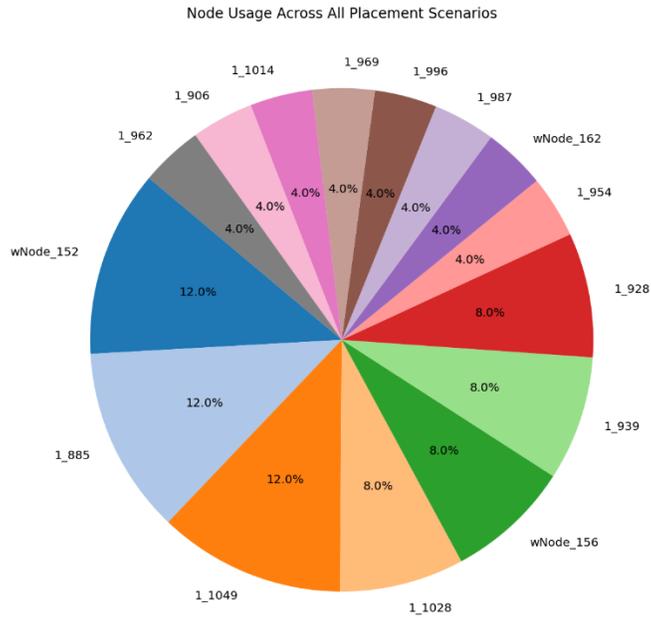

*Figure 17. Percentage of node selection in all performance objectives*

## 6.2 Scenario B: Selective contamination scenarios

In this scenario, we decided to purposely "contaminate" the distribution network, to examine the correctness of the DBPFinder under specific contamination cases. For these contamination scenarios, environmental parameters such as water temperature, total organic carbon, dissolved organic nitrogen and chlorine dosage can take a wide range of values by generating synthetically input datasets for different contamination scenarios. This means that the values can purposely be too high, exceeding the regulatory limits set, or too low, leading to very small concentrations of DBPs. In each case, by adapting the values of the environmental parameters at each node of the distribution network, a completely new contamination scenario can be generated each time. Furthermore, the injection point of some DBPs can also be randomized and may vary from one node being contaminated up to all of the nodes in the network being injected all at once. The contamination scenarios selected are defined in Table 4 below.

*Table 6. Contamination scenarios.*

| Case 1 | 20% nodes contaminated | THM |
|---|---|---|
| Case 2 | 20% nodes contaminated | HAA |
| Case 3 | 20% nodes contaminated | THM and HAA |
| Case 4 | 40% nodes contaminated | THM |
| Case 5 | 40% nodes contaminated | HAA |
| Case 6 | 40% nodes contaminated | THM and HAA |
| Case 7 | 60% nodes contaminated | THM |
| Case 8 | 60% nodes contaminated | HAA |
| Case 9 | 60% nodes contaminated | THM and HAA |
| Case 10 | 80% nodes contaminated | THM |
| Case 11 | 80% nodes contaminated | HAA |
| Case 12 | 80% nodes contaminated | THM and HAA |

For all the cases above, the performance objectives tackled are: time of detection, specific placement of DBPs for the cases of each family individually and finally, placement based on the concentration score of the nodes when multiple DBPs are present in the network. For the percentage of the nodes contaminated, the visualization of the network consists of a heatmap, that showcases the levels of concentration of the DBPs at the nodes of the network. Since this is a dead-end network, the contamination has as starting point the nodes at the center of the system and the dataset utilized follows the spatial interpolation method described in Section 3.3.

For the case of the network being 20% contaminated, the visualization is showcased in Figure 18.

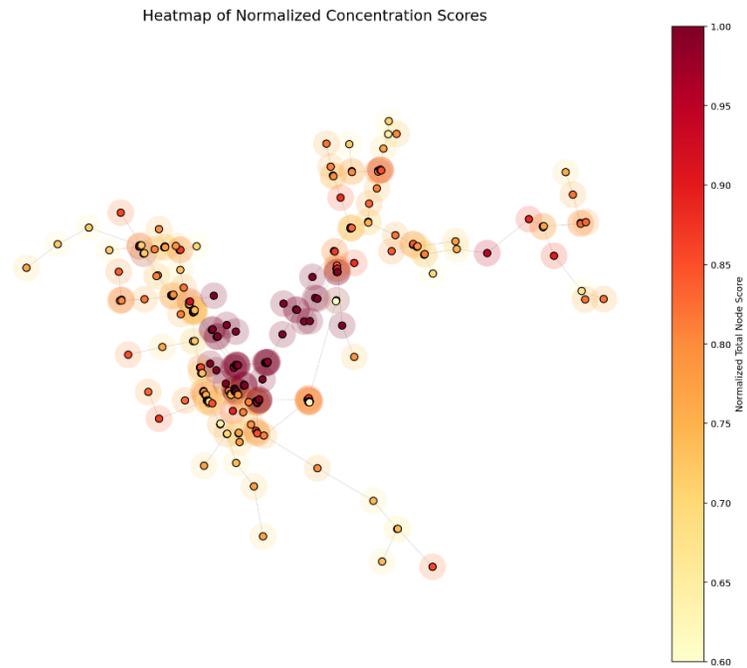

*Figure 18. 20% of the network is contaminated.*

The normalized score results are shown in Figure 19. The five nodes selected for placement of the five sensors consist of high DBP concentration of both families leading to very high normalized concentration value (above 1).

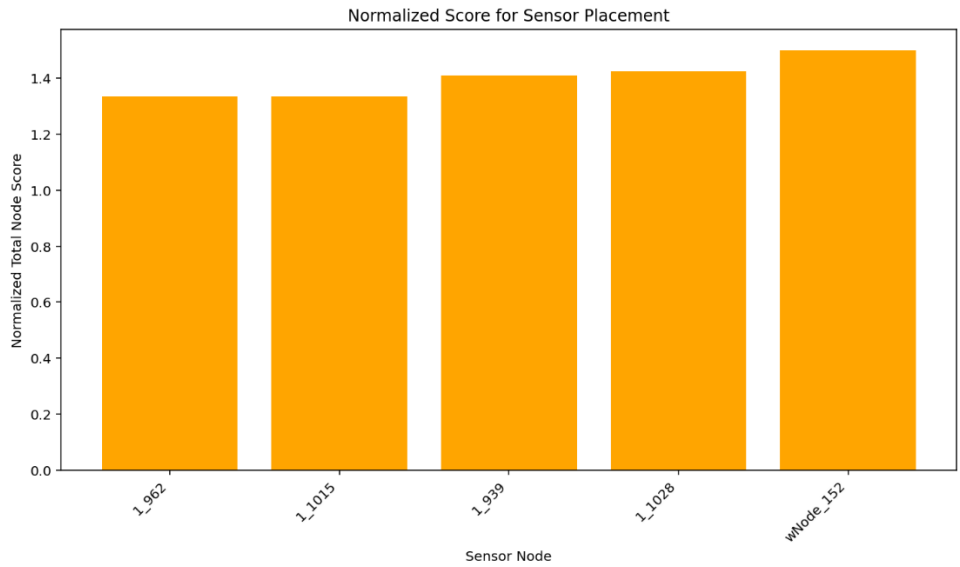

*Figure 19. Normalized score results based on five sensors.*

The network which, as described, is of a dead-end type, has specific paths that lead to the dead-end nodes. For this reason, the selected nodes for sensor placement can minimize the time of detection and cover the gaps. The time of detection for these selected nodes is seen on Figure 20.

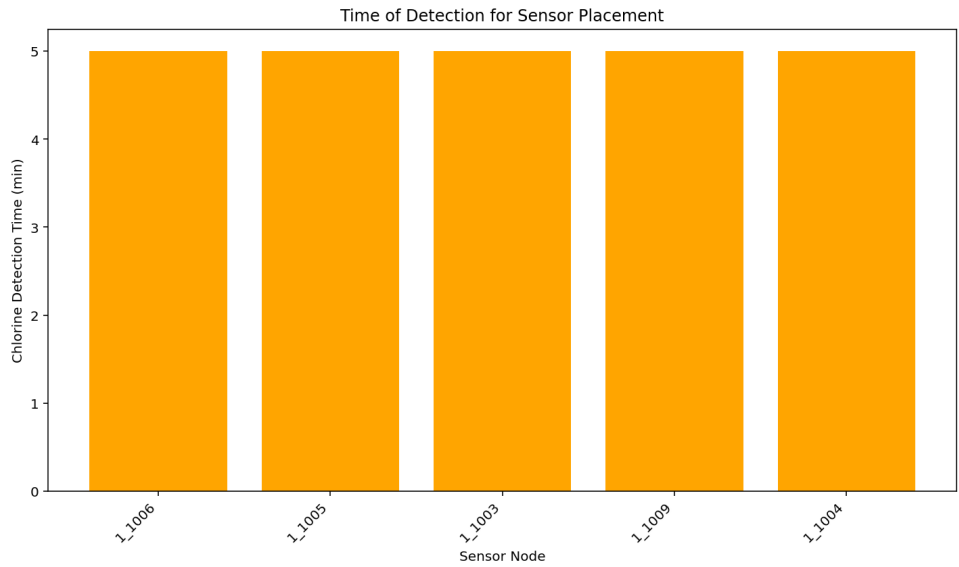

Figure 20. Time of detection for 20% contamination scenario.

Lastly, as described in Section 4.6, the events accumulated at the nodes assist in the placement of sensors when targeting specific families of DBPs. In the Figures 21 and 22, the selected nodes have the highest number of events for each family of DBPs.

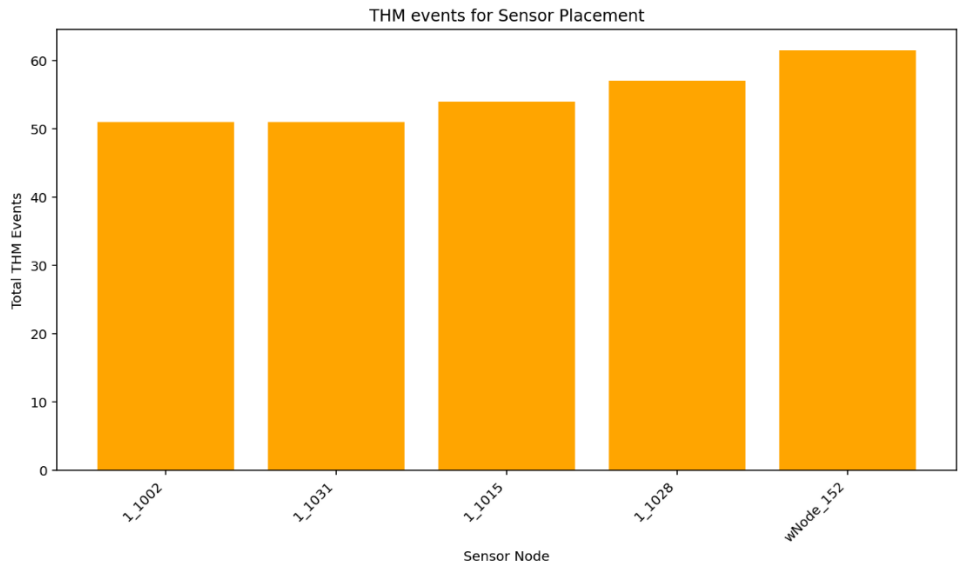

*Figure 21. Placement is based for the detection of THMs.*

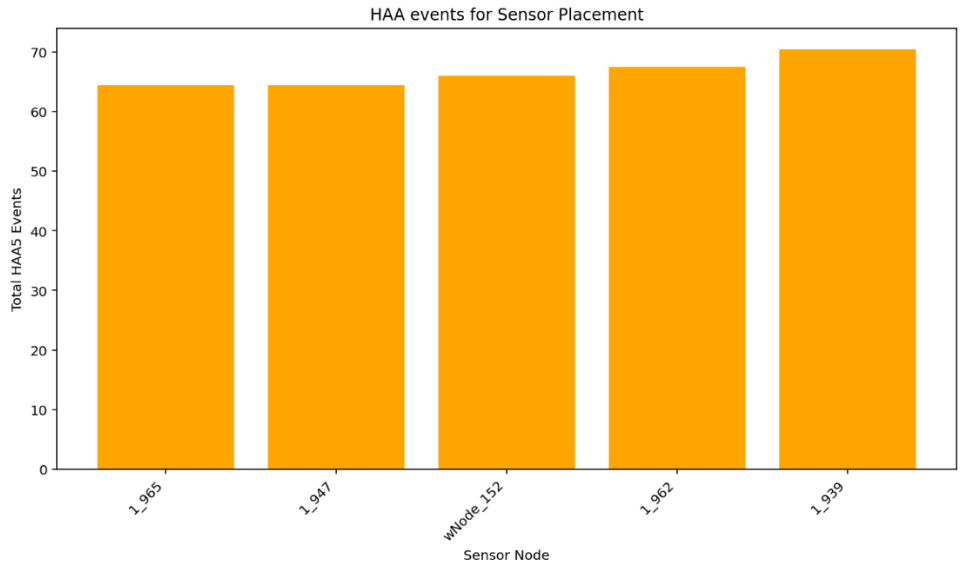

*Figure 22. Placement is based for the detection of HAAs.*

The rest of the cases of contamination in the network and their results can be found in Appendix A. In all of these contamination cases, we can observe that DBPFinder acts correctly, i.e. choosing the nodes with the highest impact in the network, based on the metrics selected, in order to maximize/minimize the value of the metric accordingly.

### 6.3  Scenario C: Scalability of DBPFinder

Regarding the scalability of the simulator, the placement of a high number of sensors was tested on two different large networks. The first network consists of roughly 500 nodes (Figure 23), being twice the size of the network of the pilot case study, which was examined in scenario A. The second network, being the largest the simulator has tested, consists of roughly 1,000 nodes (Figure 24). Both networks have been derived from the study of Bi et al.[40], in order to test DBPFinder thoroughly and were chosen due to the high number of nodes available, to allow for better stress-testing of the simulator. Table 7 showcases the time of placement for each number of sensors for the second network, which is the largest one DBPFinder has been tested on. Furthermore, the placement of the sensors considers all the performance objectives together and the processing time for the visualization outputs, leading to the most exhausting possibility for calculating placement time. In total, this includes five performance objectives being calculated at the same time (time of detection, mass consumption, concentration score, specific sensor placement for THM and HAA) and eleven visualizations of the results being processed, two figures for each objective (placement on network and bar graph) and the pie chart showcasing all the nodes that have been selected throughout the simulation.

Table 7. Time of placement for varying number of sensors.

| | |
|---|---|
| 5 sensors | 4 seconds |
| 50 sensors | 10 seconds |
| 100 sensors | 16 seconds |
| 200 sensors | 30 seconds |
| 300 sensors | 41 seconds |
| 500 sensors | 1 minute and 5 seconds |

For smaller networks, the time of placement is even lower, since the nodes that need to be filtered are even less.

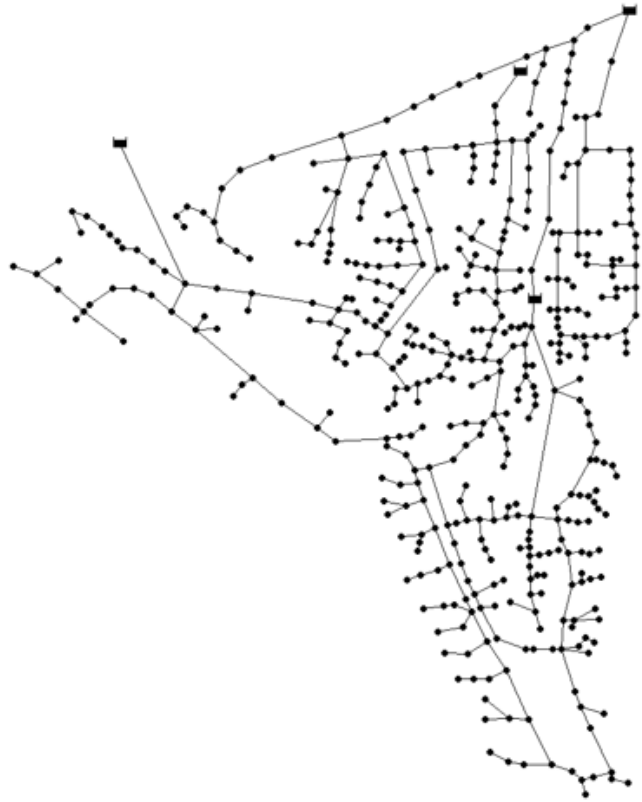

*Figure 23.* Balerma network of roughly 500 nodes.

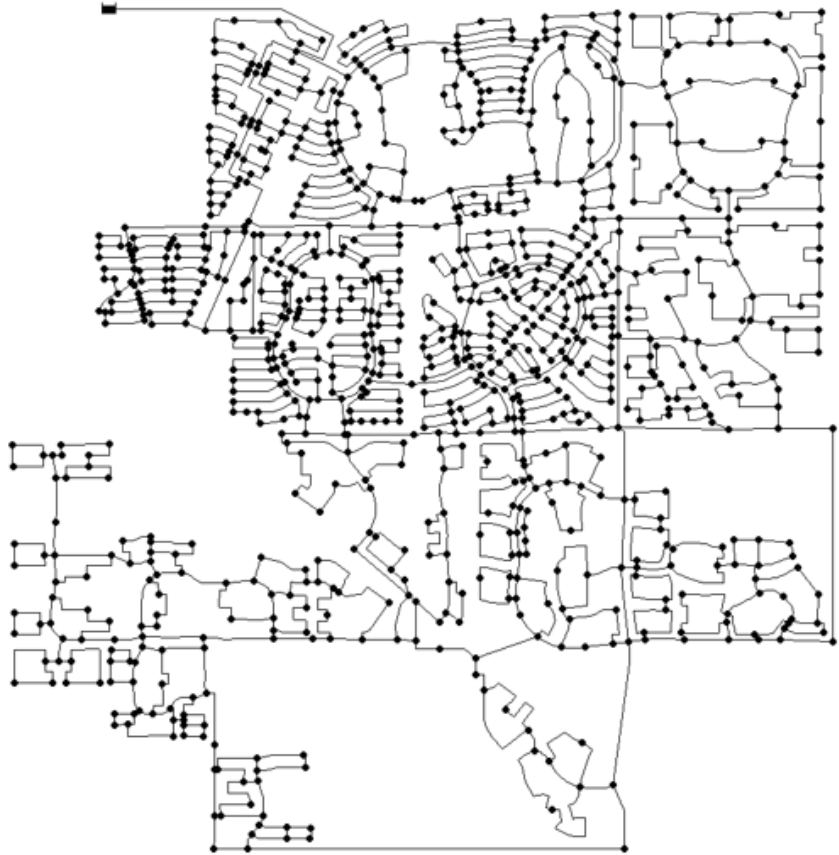

*Figure 24. KLmod network of roughly 1000 nodes.*

# 7 Discussion

The DBPFinder software aims to provide to water utility operators and researchers an important tool for performing strategic sensor placement in water distribution networks, aiming to maximize coverage as much as possible based on different scenarios and requirements of the operators. By importing relevant environmental data, the users will have access to immediate and meaningful results when it comes to optimal sensor placement, depending on the performance objectives selected. By having a list of contracts available for each node in the network, the user can also address the performance objective of mass consumption, allowing the most efficient placement to minimize the consumption of contaminants by civilians. Furthermore, the feature of adding manually equations for formation of different DBP families can provide flexibility and more options for configurations to the operators, allowing them to choose equations which have been derived in similar conditions and context of a water distribution network.

To handle multiple performance objectives simultaneously, the software employs an exhaustive approach, systematically evaluating all candidate locations for sensor placement. It selects the most suitable nodes based on each performance objective, progressively narrowing down the options according to DBP concentration levels. This method reduces the likelihood of suboptimal placements and increases the robustness of the final output, depending on the needs of the water utility operators each time. Ultimately, the physical installation of sensors is carried out by water utility operators, who must ensure that selected nodes have the necessary infrastructure, such as access to electricity and water and that the locations are safe and suitable for the equipment.

Moreover, DBPFinder provides high transparency, allowing to understand the reasoning behind certain placement recommendation, as it provides complete calculations as performed at the individual nodes of the network.

Finally, the DBPFinder tool allows to understand how many sensors are needed to reach certain performance based on the metrics considered and help the operators take more informed decisions in terms of purchase of equipment that satisfies their needs.

## 7.1 Limitations

As previously mentioned, one of the biggest limitations is the lack of data, which makes it difficult to construct a complete scenario of strategic sensor placement having access to all parameters affecting such a decision. Thus, generating synthetic data was the only feasible way to approach the partial data available, filling the gaps in a way that was as meaningful and realistic as possible. Furthermore, the lack of publicly available real-world calibrated water distribution networks was an issue that limited the validation efforts of the DBPFinder tool, forcing us to articulate our own water networks and to perform hypothetical scenarios.

With respect to the concentration of DBPs at each node, the equations that are imported for our simulation software are extracted from empirical studies and from lab experiments based on varying water distribution networks from all over the world, with different characteristics, perhaps not generalizing enough for all networks, perhaps not reflecting the functional groups present in natural organic matter unique to some specific water network. As a result, achieving realistic and precise concentration values is only possible through the analysis of water samples from some specific network of interest, followed by the development of tailored equations based on experimental results.

In the simulations conducted for strategic sensor placement in Section 6, sensors were assumed to be perfect, an assumption that does not align with real-world conditions. In practice, sensors can malfunction for various reasons, such as data storage limitations, electrical failures, water quality issues, or even vandalism, particularly in accessible public locations. At the same time, sensors might not be fully calibrated or accurate, having erroneous measurements which reduce overall accuracy of measurements. Simulating such errors is not currently feasible within the scope of this paper, as these events are inherently random and highly context dependent.

### 7.2 Open-source Software

Since this software has been already tested in a real-life scenario with a calibrated water distribution network (Coimbra, Portugal), an important aim was to release an open-source version of this software that allows water utility operators and researchers to import their own data and water distribution networks and identify the best possible sensor locations for the performance objectives selected. This software guides the user through examples on how to provide clear and concise data, and allows the selection of different equations extracted from empirical studies or input their own formula. Furthermore, performance objectives can be selected, either through a single-objective or a multi-objective approach and the priority of each disinfection byproduct can be selected. This means that the placement can vary depending on which family of DBP has a higher weight than the over. The open-source version of this software can be found in the following GitHub link: https://github.com/superworld-cyens/DBPFinder.

### 7.3 Future work

The DBPFinder simulator continues to be validated in the real-world setting of the actual water distribution network of Coimbra, Portugal, based on 2 analyzers serving as ground truth data and 5 sensors acting as the low-end ones, being recommended for placement around the network on a weekly basis. This pilot is part of the H2OFORALL European project, and the pilot runs for six months. Future work includes more validation using more diverse scenarios and larger number of DBP families, while the possibility to target and prioritize individual DBP compounds will be considered. New features may be added to the DBPFinder simulator, after further consultation with beneficiaries and end-users. We welcome other operators of WDN around Europe to use DBPFinder and help us better shape the tool towards their needs, to achieve the development of a generic, accepted and relevant tool for strategic sensor placement with applicability all around the world.

## 8 Conclusion

In this work, we proposed a methodology and a simulation software for strategic placement of sensors, aiming to optimize the detection of dissinfection byproducts (DBPs), formed when chlorinated water interacts with natural occurring matter. DBPFinder, presented in this paper, is an open-source web-based tool designed to perform strategic sensor placement in chlorinated water distribution networks, Developed with Streamlit using EPANET hydraulic models, it integrates environmental data and customizable DBP predictive equations, in order to identify the most effective sensor locations for monitoring compounds like trihalomethanes (THMs) and haloacetic acids (HAAs). The tool considers different performance objectives, such as minimizing time of detection, minimizing mass consumption, plus  optimally placing sensors after prioritizing detection of specific DBP families. DBPFinder is highly transparent in the sense that it visualizes placements and scores calculated at different nodes of the network, which affect the placement decisions. Certain limitations are discussed, together with current and future research directions.

## 9 Acknowledgements

Funded by the European Union, under the Grant Agreement GA101081953 attributed to the project H2OforAll Innovative Integrated Tools and Technologies to Protect and Treat Drinking Water from DBPs. Views and opinions expressed are, however, those of the author(s) only and do not necessarily reflect those of the European Union. Neither the European Union nor the granting authority can be held responsible for them.

## 10 References

[1]     L. Lin, H. Yang, and X. Xu, "Effects of Water Pollution on Human Health and Disease Heterogeneity: A Review," *Front. Environ. Sci.*, vol. 10, p. 880246, June 2022, doi: 10.3389/fenvs.2022.880246.

[2]     S. E. Barrett, S. W. Krasner, G. L. Amy, American Chemical Society, and American Chemical Society, Eds., *Natural organic matter and disinfection by-products characterization and control in drinking water*. in ACS symposium series, no. 761. Washington, D.C: American Chemical Society, 2000.

[3]     J. D. Ritchie and E. M. Perdue, "Proton-binding study of standard and reference fulvic acids, humic acids, and natural organic matter," *Geochim. Cosmochim. Acta*, vol. 67, no. 1, pp. 85–96, Jan. 2003, doi: 10.1016/S0016-7037(02)01044-X.

[4]     M. Yan, G. V. Korshin, and H.-S. Chang, "Examination of disinfection by-product (DBP) formation in source waters: A study using log-transformed differential spectra," *Water Res.*, vol. 50, pp. 179–188, Mar. 2014, doi: 10.1016/j.watres.2013.11.028.

[5]     J. Raich and European Commission, Eds., *Review of sensors to monitor water quality: ERNCIP thematic area "Chemical & biological risks in the water sector", deliverable D1 - task 1*. Luxembourg: Publications Office, 2013. doi: 10.2788/35499.

[6]     L. Wang *et al.*, "Health risk assessment via ingestion of disinfection by-products in drinking water," *Sci. Rep.*, vol. 15, no. 1, Jan. 2025, doi: 10.1038/s41598-024-84094-9.

[7]     I. Kalita, A. Kamilaris, P. Havinga, and I. Reva, "Assessing the Health Impact of Disinfection Byproducts in Drinking Water," *ACS EST Water*, vol. 4, no. 4, pp. 1564–1578, Apr. 2024, doi: 10.1021/acsestwater.3c00664.

[8]     A. Magklis and A. Kamilaris, *DBPFinder*. [Online]. Available: https://github.com/superworld-cyens/DBPFinder

[9]     G. A. Boorman, "Drinking water disinfection byproducts: review and approach to toxicity evaluation.," *Environ. Health Perspect.*, vol. 107, no. suppl 1, pp. 207–217, Feb. 1999, doi: 10.1289/ehp.99107s1207.

[10]    R. D. Morris, A. M. Audet, I. F. Angelillo, T. C. Chalmers, and F. Mosteller, "Chlorination, chlorination by-products, and cancer: a meta-analysis.," *Am. J. Public Health*, vol. 82, no. 7, pp. 955–963, July 1992, doi: 10.2105/AJPH.82.7.955.

[11]    K. Yeung, L. Xie, P. Nair, and H. Peng, "Haloacetonitriles Induce Structure-Related Cellular Toxicity Through Distinct Proteome Thiol Reaction Mechanisms," *ACS Environ. Au*, vol. 5, no. 1, pp. 101–113, Jan. 2025, doi: 10.1021/acsenvironau.4c00068.

[12]    J. Fu, W.-N. Lee, C. Coleman, K. Nowack, J. Carter, and C.-H. Huang, "Removal of disinfection byproduct (DBP) precursors in water by two-stage biofiltration treatment," *Water Res.*, vol. 123, pp. 224–235, Oct. 2017, doi: 10.1016/j.watres.2017.06.073.

[13]    R. Sadiq and M. Rodriguez, "Disinfection by-products (DBPs) in drinking water and predictive models for their occurrence: a review," *Sci. Total Environ.*, vol. 321, no. 1–3, pp. 21–46, Apr. 2004, doi: 10.1016/j.scitotenv.2003.05.001.

[14]     Y. Khor, A. R. A. Aziz, and S. S. Chong, "Recent developments and sustainability in monitoring chlorine residuals for water quality control: a critical review," *RSC Sustain.*, vol. 2, no. 9, pp. 2468–2485, 2024, doi: 10.1039/D4SU00188E.

[15]     M. H. Banna, H. Najjaran, R. Sadiq, S. A. Imran, M. J. Rodriguez, and M. Hoorfar, "Miniaturized water quality monitoring pH and conductivity sensors," *Sens. Actuators B Chem.*, vol. 193, pp. 434–441, Mar. 2014, doi: 10.1016/j.snb.2013.12.002.

[16]     Y. Qin *et al.*, "Integrated water quality monitoring system with pH, free chlorine, and temperature sensors," *Sens. Actuators B Chem.*, vol. 255, pp. 781–790, Feb. 2018, doi: 10.1016/j.snb.2017.07.188.

[17]     P. M. Ramos, J. M. D. Pereira, H. M. G. Ramos, and A. L. Ribeiro, "A Four-Terminal Water-Quality-Monitoring Conductivity Sensor," *IEEE Trans. Instrum. Meas.*, vol. 57, no. 3, pp. 577–583, Mar. 2008, doi: 10.1109/TIM.2007.911703.

[18]     D. Gillett and A. Marchiori, "A Low-Cost Continuous Turbidity Monitor," *Sensors*, vol. 19, no. 14, p. 3039, July 2019, doi: 10.3390/s19143039.

[19]     E. T. De Camargo *et al.*, "Low-Cost Water Quality Sensors for IoT: A Systematic Review," *Sensors*, vol. 23, no. 9, p. 4424, Apr. 2023, doi: 10.3390/s23094424.

[20]     C. Hu, M. Li, D. Zeng, and S. Guo, "A survey on sensor placement for contamination detection in water distribution systems," *Wirel. Netw.*, vol. 24, no. 2, pp. 647–661, Feb. 2018, doi: 10.1007/s11276-016-1358-0.

[21]     M. Weickgenannt, Z. Kapelan, M. Blokker, and D. A. Savic, "Risk-Based Sensor Placement for Contaminant Detection in Water Distribution Systems," *J. Water Resour. Plan. Manag.*, vol. 136, no. 6, pp. 629–636, Nov. 2010, doi: 10.1061/(ASCE)WR.1943-5452.0000073.

[22]     Z. Ohar, O. Lahav, and A. Ostfeld, "Optimal sensor placement for detecting organophosphate intrusions into water distribution systems," *Water Res.*, vol. 73, pp. 193–203, Apr. 2015, doi: 10.1016/j.watres.2015.01.024.

[23]     S. Sophocleous, D. A. Savić, Z. Kapelan, and O. Giustolisi, "A Two-stage Calibration for Detection of Leakage Hotspots in a Real Water Distribution Network," *Procedia Eng.*, vol. 186, pp. 168–176, 2017, doi: 10.1016/j.proeng.2017.03.223.

[24]     K. Liu, T. Lin, T. Zhong, X. Ge, F. Jiang, and X. Zhang, "New methods based on a genetic algorithm back propagation (GABP) neural network and general regression

neural network (GRNN) for predicting the occurrence of trihalomethanes in tap water," *Sci. Total Environ.*, vol. 870, p. 161976, Apr. 2023, doi: 10.1016/j.scitotenv.2023.161976.

[25]     B. Dong, S. Shu, and D. Li, "Optimization of Secondary Chlorination in Water Distribution Systems for Enhanced Disinfection and Reduced Chlorine Odor Using Deep Belief Network and NSGA-II," *Water*, vol. 16, no. 18, p. 2666, Sept. 2024, doi: 10.3390/w16182666.

[26]     J. Berry, W. E. Hart, C. A. Phillips, J. G. Uber, and J.-P. Watson, "Sensor Placement in Municipal Water Networks with Temporal Integer Programming Models," *J. Water Resour. Plan. Manag.*, vol. 132, no. 4, pp. 218–224, July 2006, doi: 10.1061/(ASCE)0733-9496(2006)132:4(218).

[27]     D. L. Boccelli and W. E. Hart, "Optimal Monitoring Location Selection for Water Quality Issues," in *World Environmental and Water Resources Congress 2007*, Tampa, Florida, United States: American Society of Civil Engineers, May 2007, pp. 1–8. doi: 10.1061/40927(243)522.

[28]     I. Delpla, M. Florea, G. Pelletier, and M. J. Rodriguez, "Optimizing disinfection by-product monitoring points in a distribution system using cluster analysis," *Chemosphere*, vol. 208, pp. 512–521, Oct. 2018, doi: 10.1016/j.chemosphere.2018.06.009.

[29]     A. Ardila, M. J. Rodriguez, and G. Pelletier, "Optimizing sampling location for water quality degradation monitoring in distribution systems: Assessing global representativeness and potential health risk," *J. Environ. Manage.*, vol. 365, p. 121505, Aug. 2024, doi: 10.1016/j.jenvman.2024.121505.

[30]     A. K. Salem and A. A. Abokifa, "Optimal Sensor Placement in Water Distribution Networks Using Dynamic Prediction Graph Neural Networks," in *The 3rd International Joint Conference on Water Distribution Systems Analysis & Computing and Control for the Water Industry (WDSA/CCWI 2024)*, MDPI, Sept. 2024, p. 171. doi: 10.3390/engproc2024069171.

[31]     R. Łangowski and M. A. Brdys, "An optimised placement of the hard quality sensors for a robust monitoring of the chlorine concentration in drinking water distribution systems," *J. Process Control*, vol. 68, pp. 52–63, Aug. 2018, doi: 10.1016/j.jprocont.2018.04.007.

[32]     E. Q. Shahra and W. Wu, "Water contaminants detection using sensor placement approach in smart water networks," *J. Ambient Intell. Humaniz. Comput.*, vol. 14, no. 5, pp. 4971–4986, May 2023, doi: 10.1007/s12652-020-02262-x.

[33]     C. M. Villanueva *et al.*, "Global assessment of chemical quality of drinking water: The case of trihalomethanes," *Water Res.*, vol. 230, p. 119568, Feb. 2023, doi: 10.1016/j.watres.2023.119568.

[34]     M. Malika, B. Hassiba, L. Radouane, B. Nabil, B. Samir, and M. Trari, "Chlorine Decay Simulation in Real Water Distribution System Using Epanet Program," *Ann. West Univ. Timisoara - Phys.*, vol. 65, no. 1, pp. 80–96, Dec. 2023, doi: 10.2478/awutp-2023-0007.

[35]     S. Chowdhury, P. Champagne, and P. J. McLellan, "Models for predicting disinfection byproduct (DBP) formation in drinking waters: A chronological review," *Sci. Total Environ.*, vol. 407, no. 14, pp. 4189–4206, July 2009, doi: 10.1016/j.scitotenv.2009.04.006.

[36]     V. Uyak, K. Ozdemir, and I. Toroz, "Multiple linear regression modeling of disinfection by-products formation in Istanbul drinking water reservoirs," *Sci. Total Environ.*, vol. 378, no. 3, pp. 269–280, June 2007, doi: 10.1016/j.scitotenv.2007.02.041.

[37]     A. I. Okoji, C. N. Okoji, and O. S. Awarun, "Performance evaluation of artificial intelligence with particle swarm optimization (PSO) to predict treatment water plant DBPs (haloacetic acids)," *Chemosphere*, vol. 344, p. 140238, Dec. 2023, doi: 10.1016/j.chemosphere.2023.140238.

[38]     H. Hong *et al.*, "Using regression models to evaluate the formation of trihalomethanes and haloacetonitriles via chlorination of source water with low SUVA values in the Yangtze River Delta region, China," *Environ. Geochem. Health*, vol. 38, no. 6, pp. 1303–1312, Dec. 2016, doi: 10.1007/s10653-016-9797-1.

[39]     J. Sohn and H.-S. Kim, "Modeling of chlorination disinfection by-products (DBPs) formation at the full-scale study," *KSCE J. Civ. Eng.*, vol. 6, no. 1, pp. 89–100, Mar. 2002, doi: 10.1007/BF02829039.

[40]     W. Bi, G. C. Dandy, and H. R. Maier, "Improved genetic algorithm optimization of water distribution system design by incorporating domain knowledge," *Environ. Model. Softw.*, vol. 69, pp. 370–381, July 2015, doi: 10.1016/j.envsoft.2014.09.010.

# Appendix A

In this Appendix, the rest of the visualizations of Scenario B from Section 6.2 are presented. In the Figures below the results of 20 sensors being placed on the 20% contaminated network are shown.

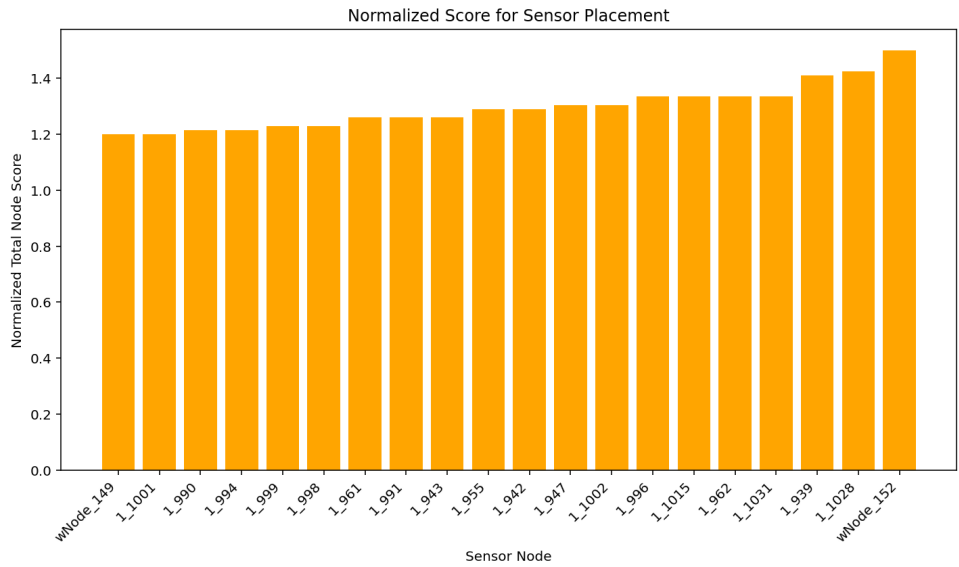

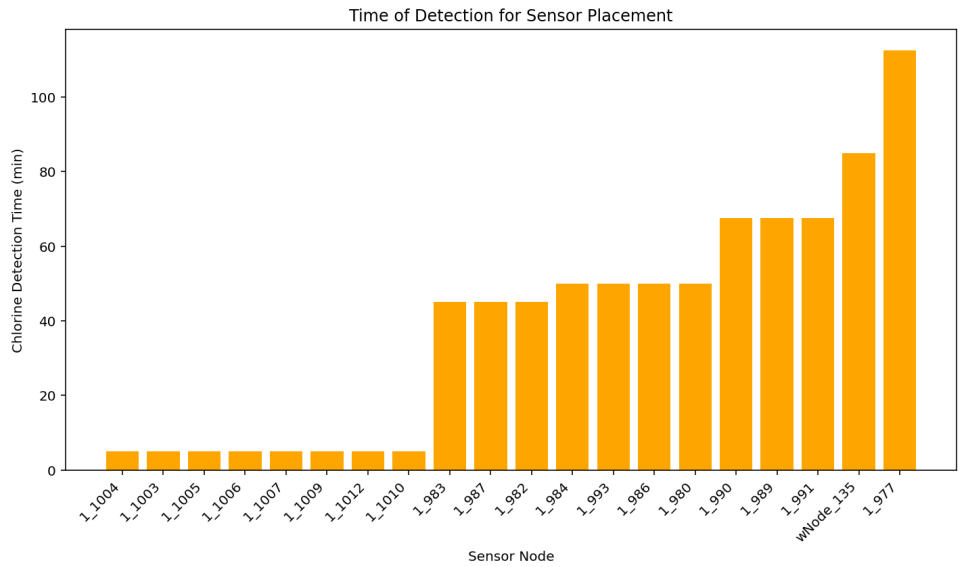
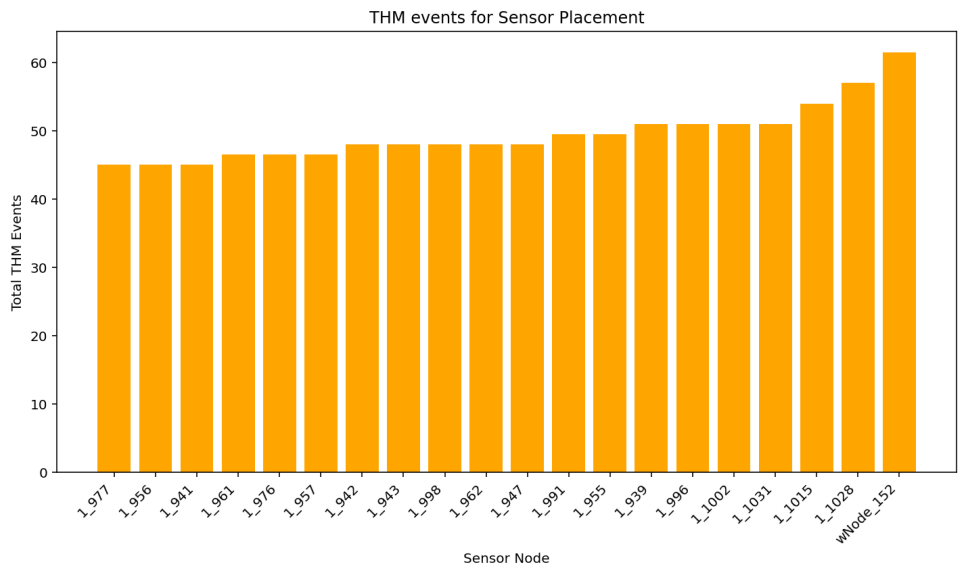

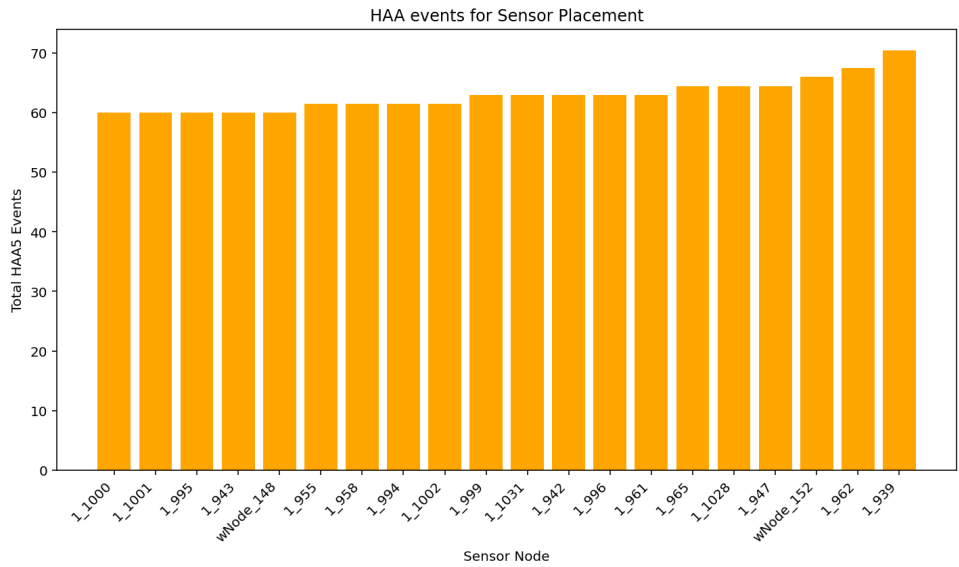

For the next case of contamination, the network is infected to up to 40% of the nodes. The heatmap of the network and the results are presented below (for 5 and 20 sensors placed).

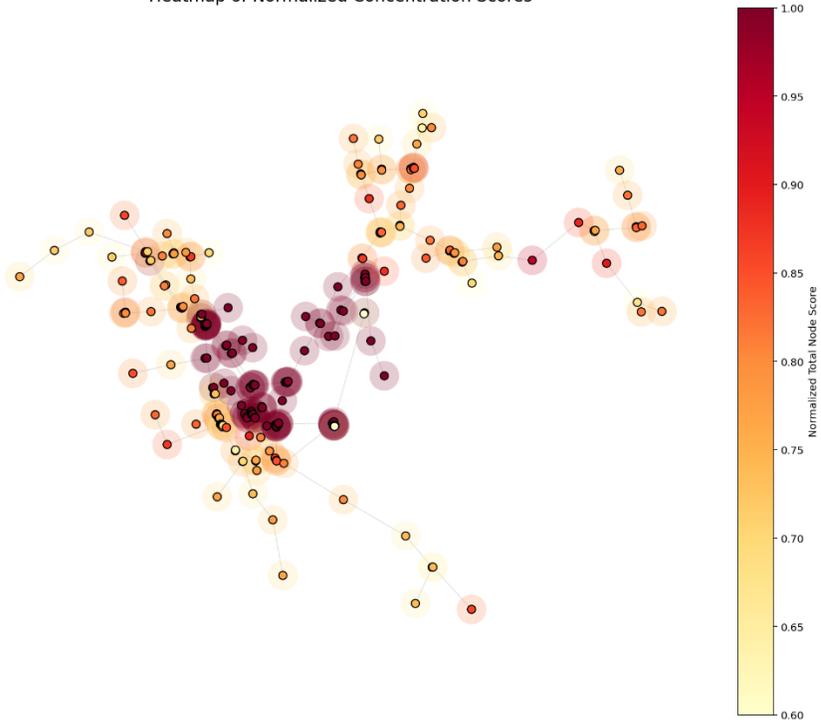

Heatmap of Normalized Concentration Scores

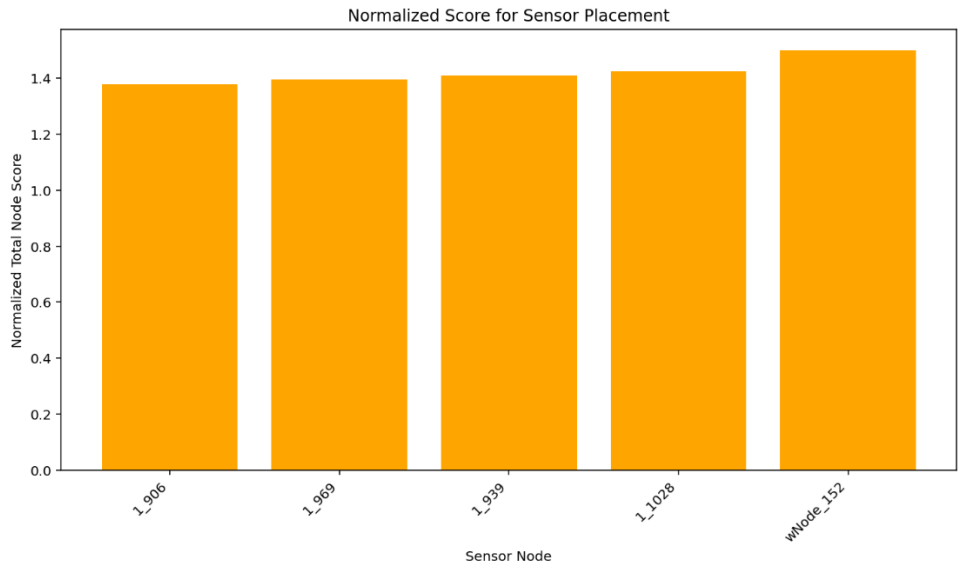
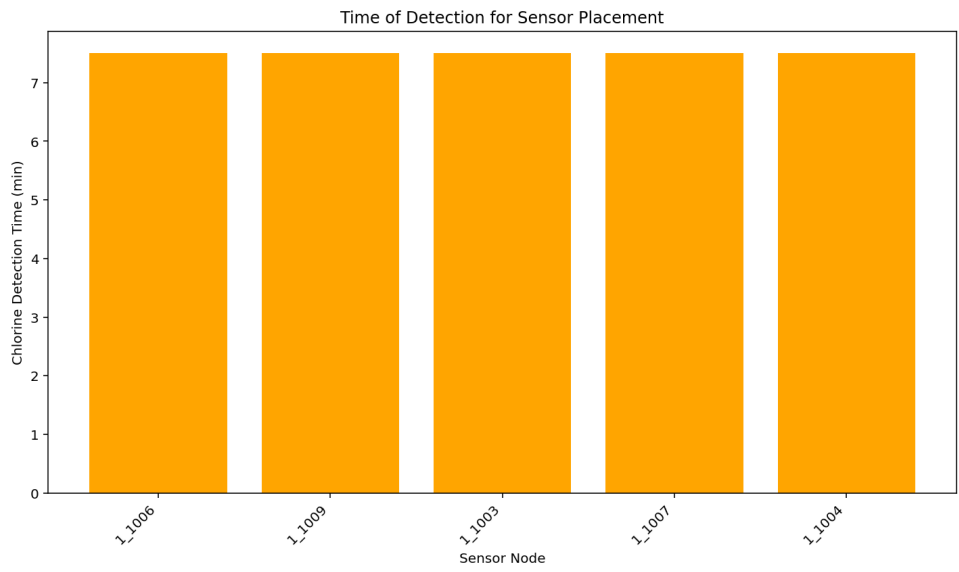

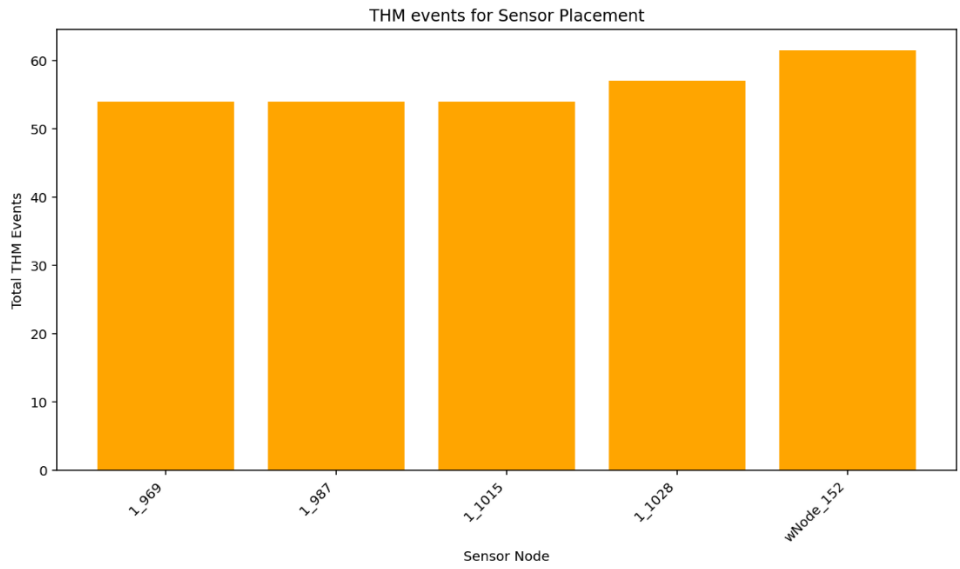
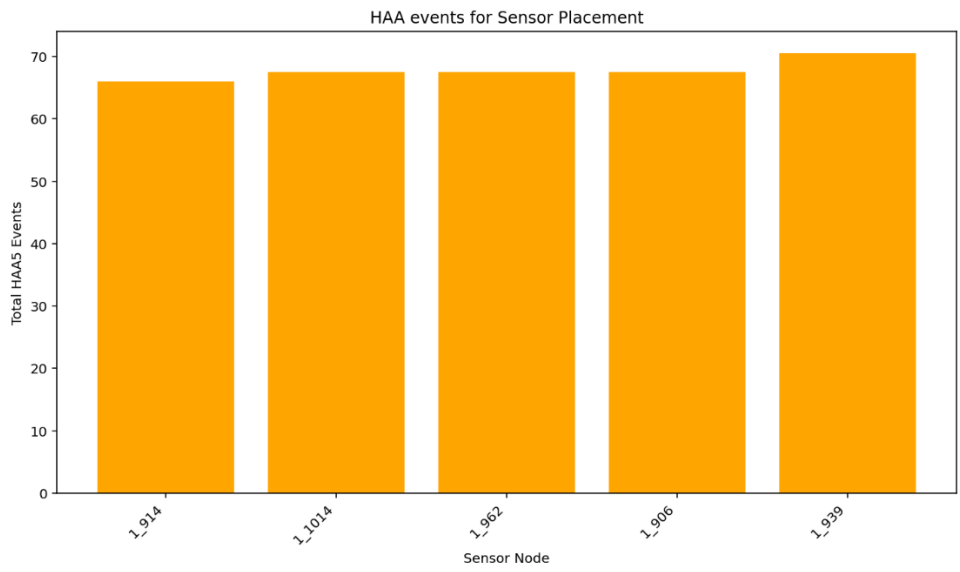

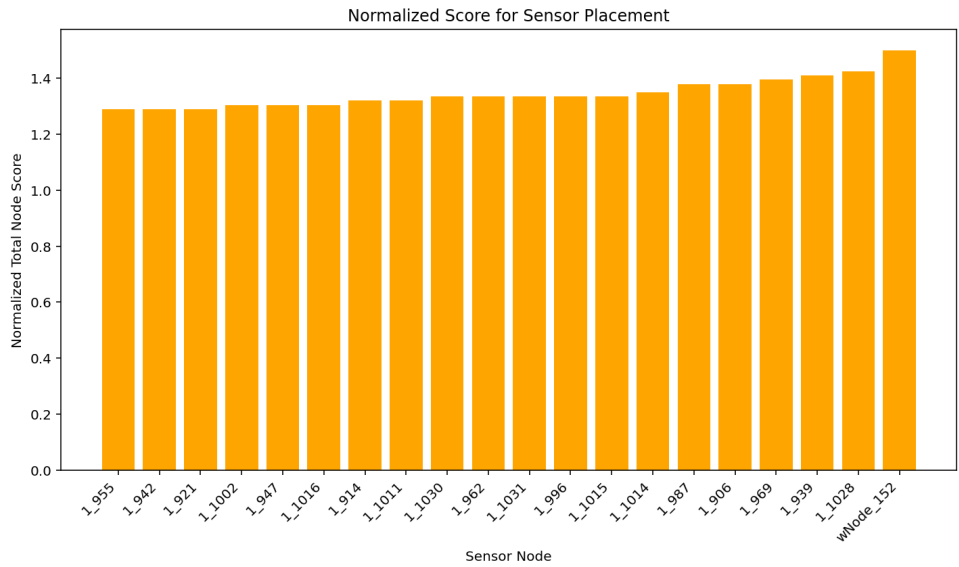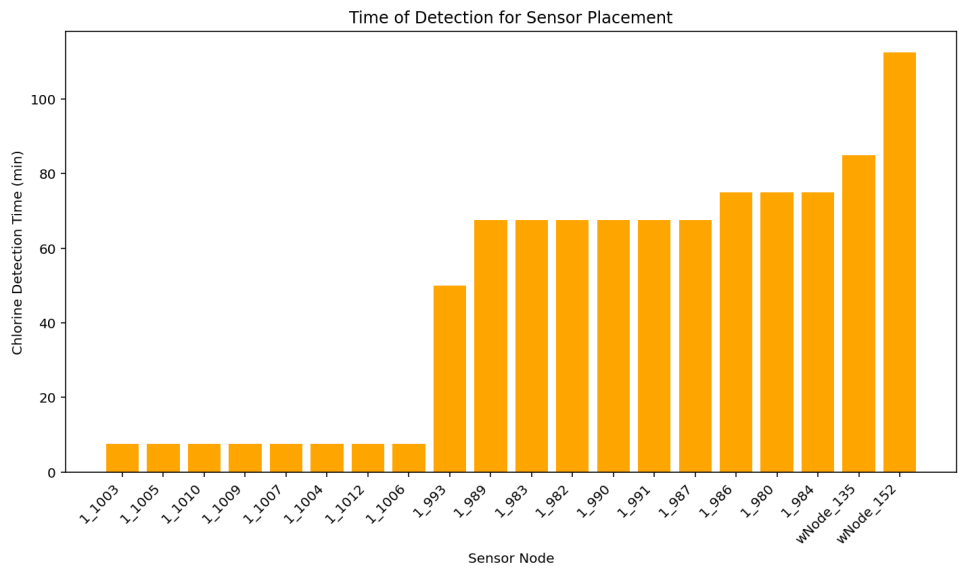

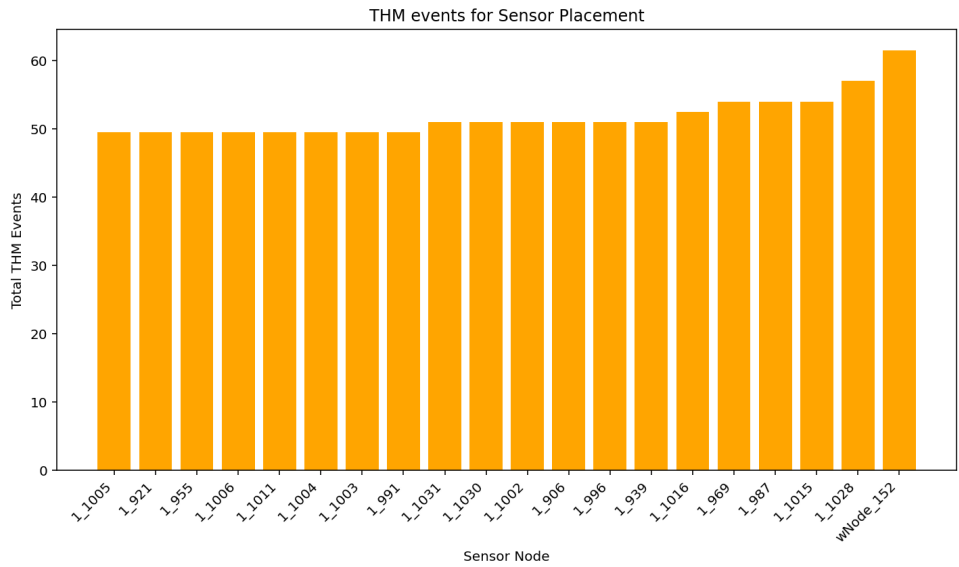

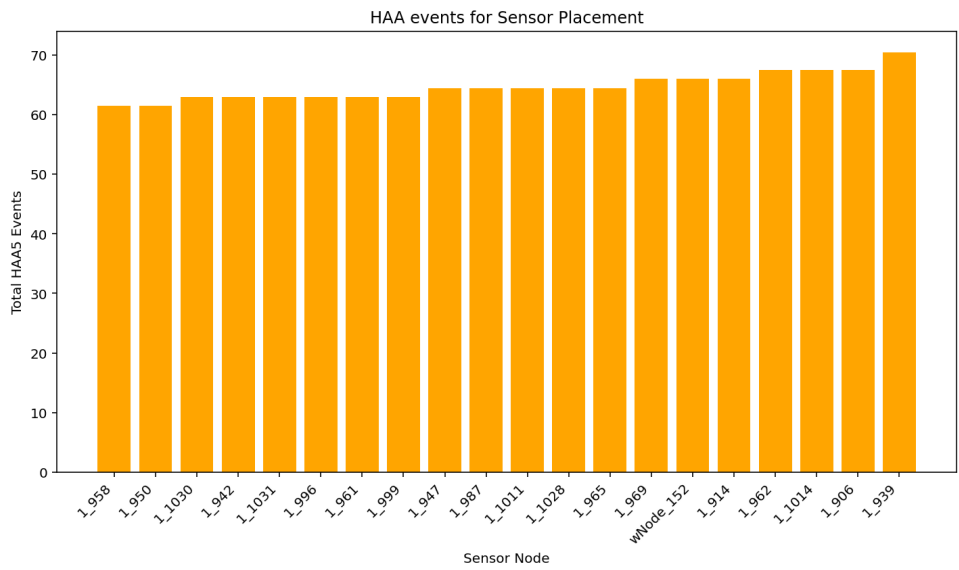

For the third case of contamination of the network, the network is polluted to up to 60% of the nodes. The results are showcased below.

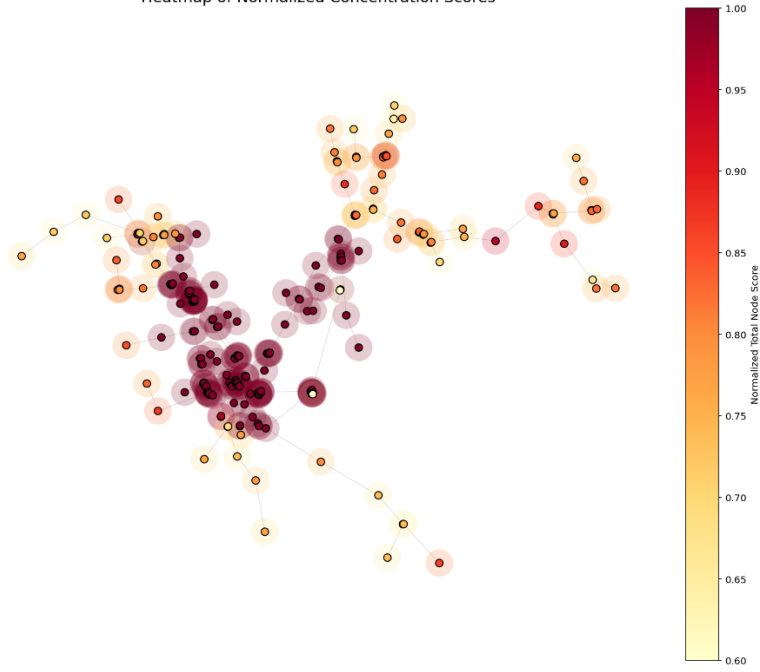

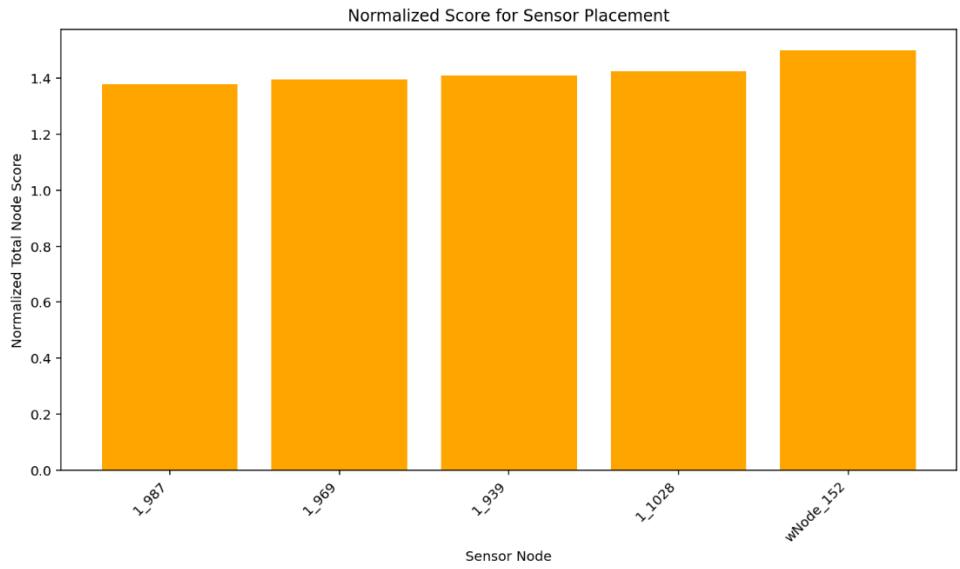
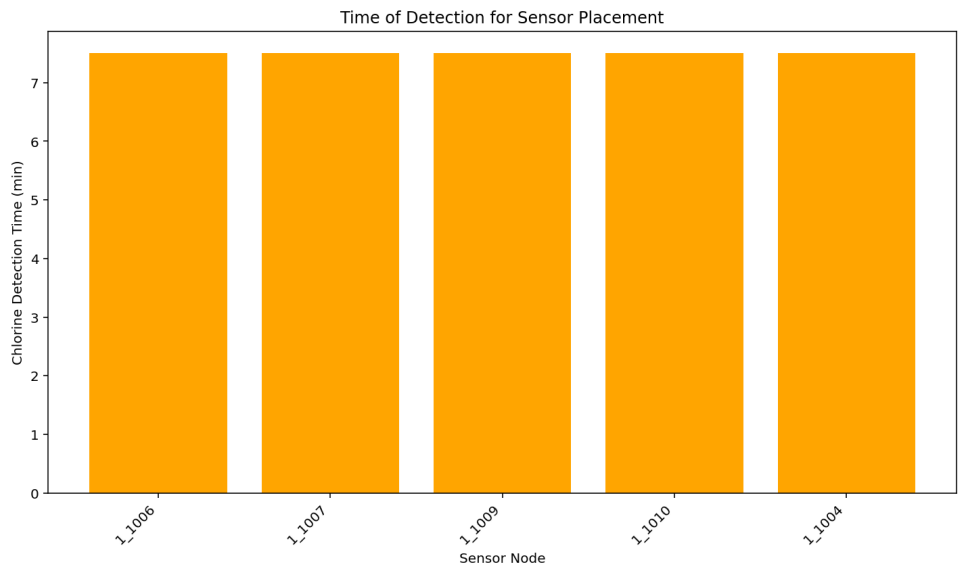

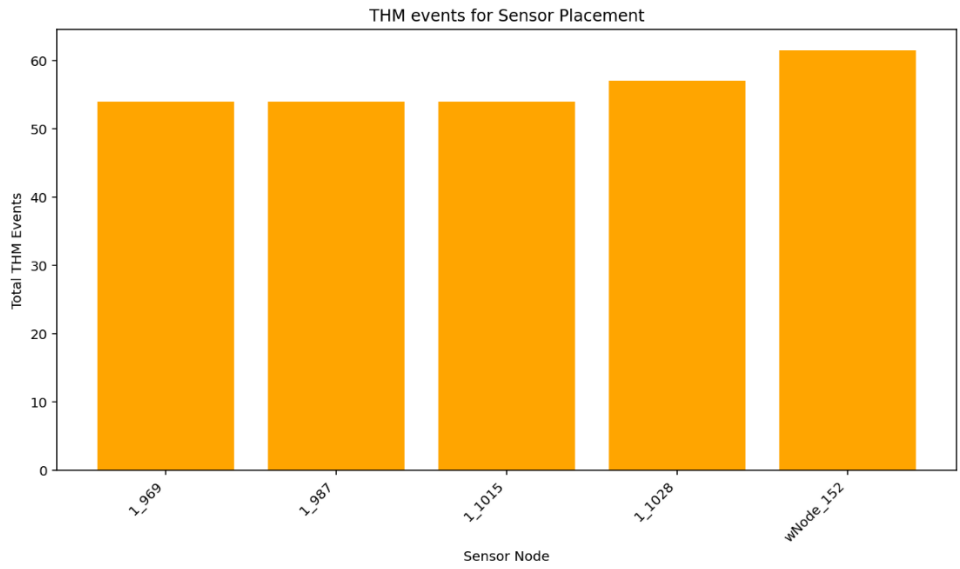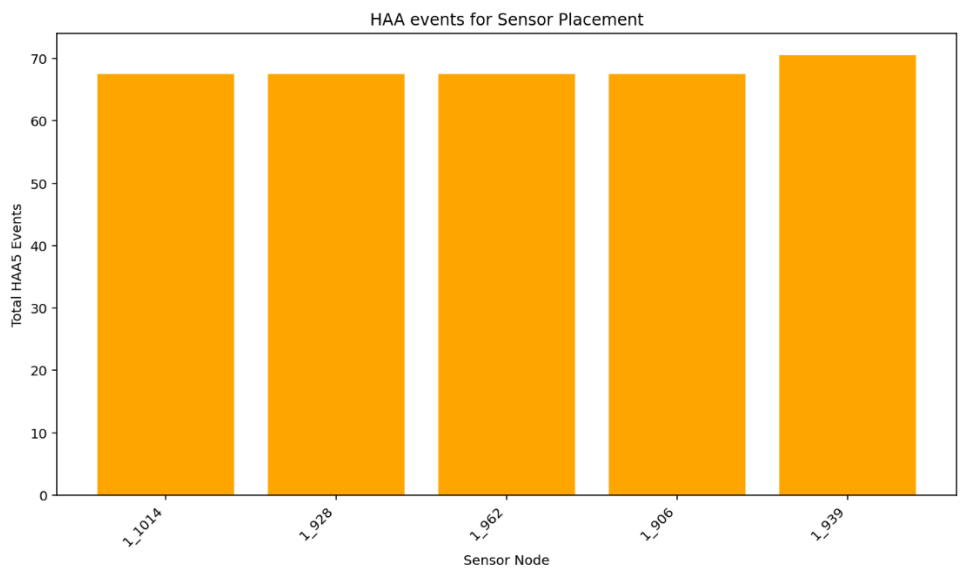

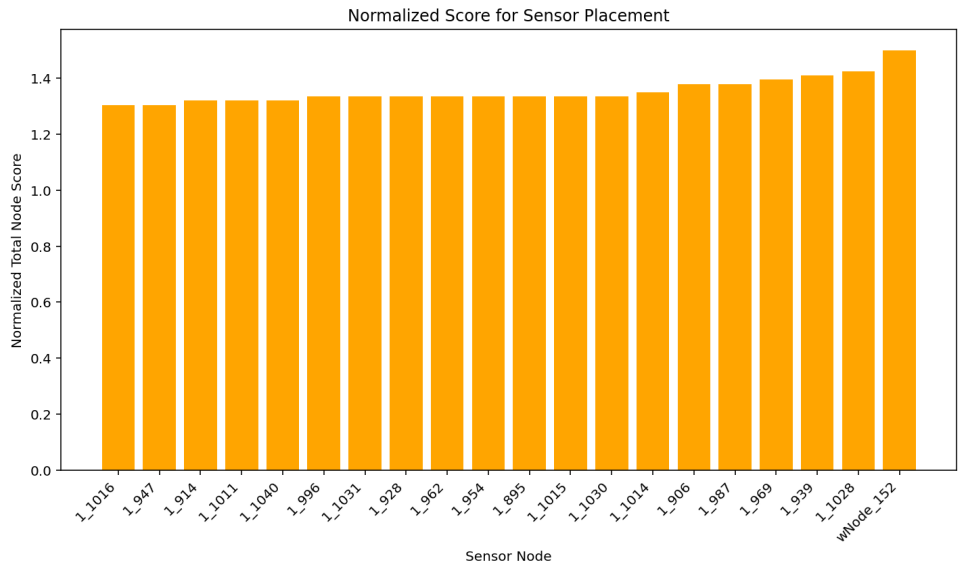
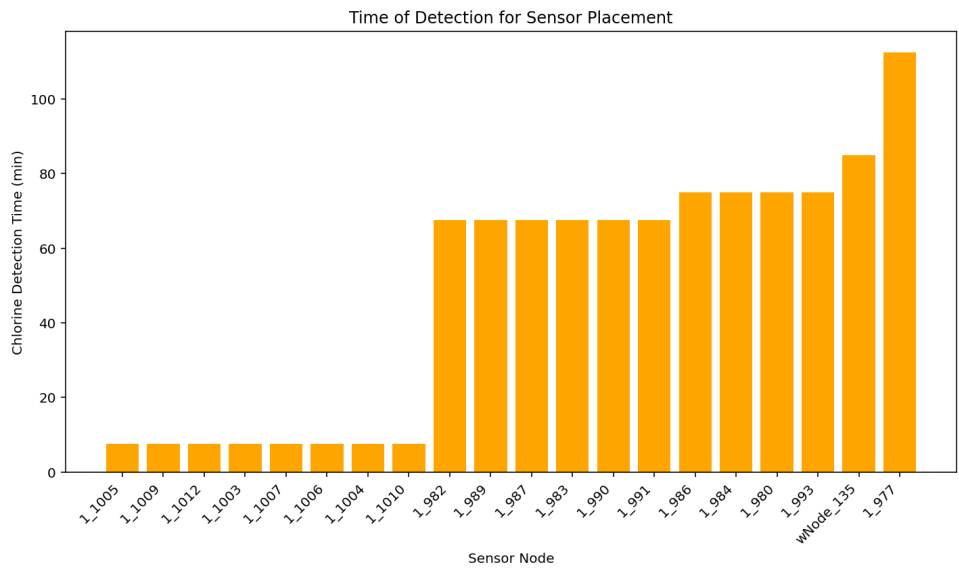

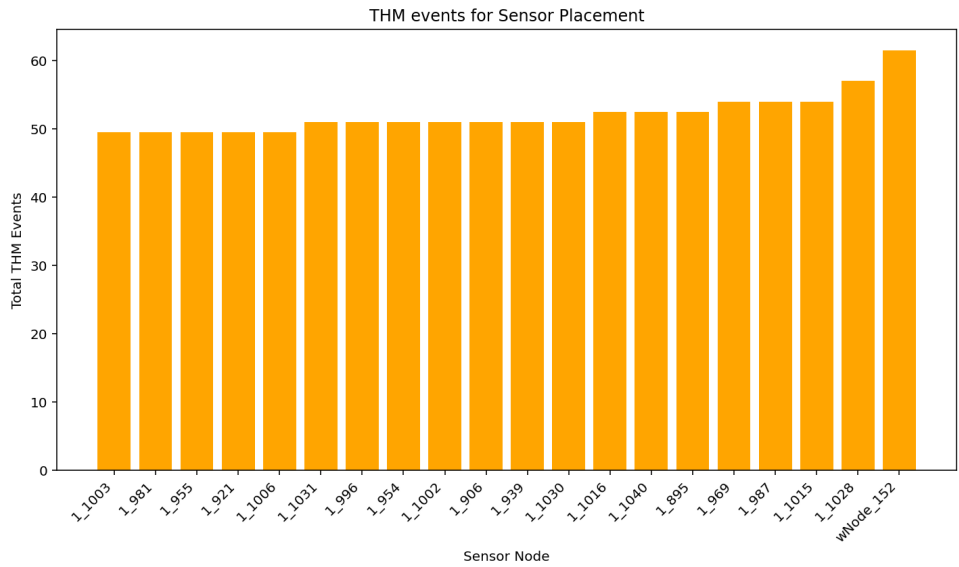

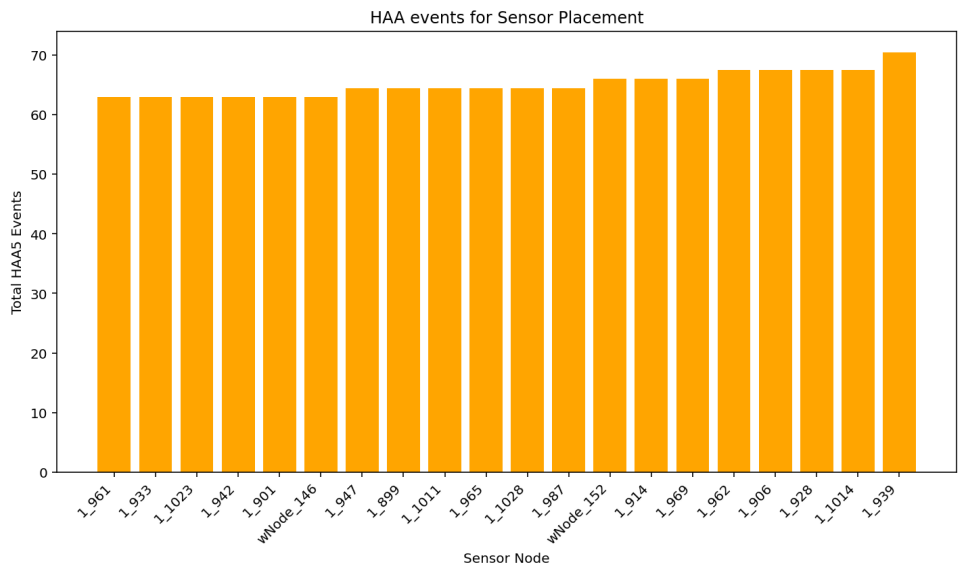

The last case for Scenario B includes a contaminated network of up to 80% of the nodes. The results are shown in the Figures below.

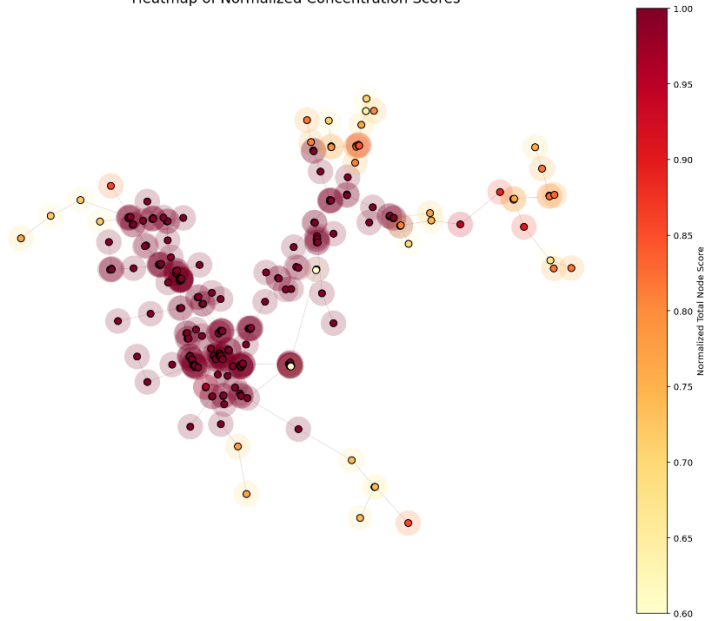

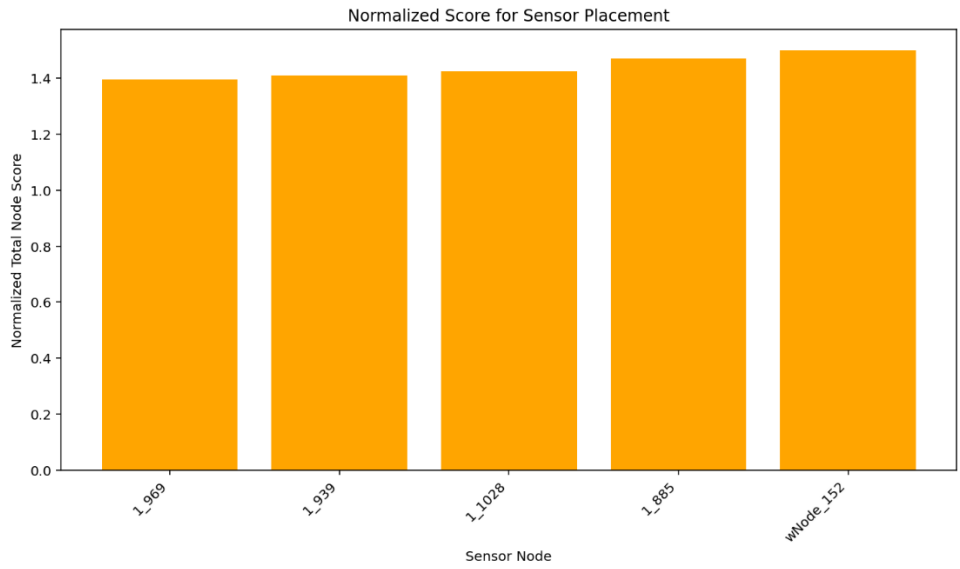
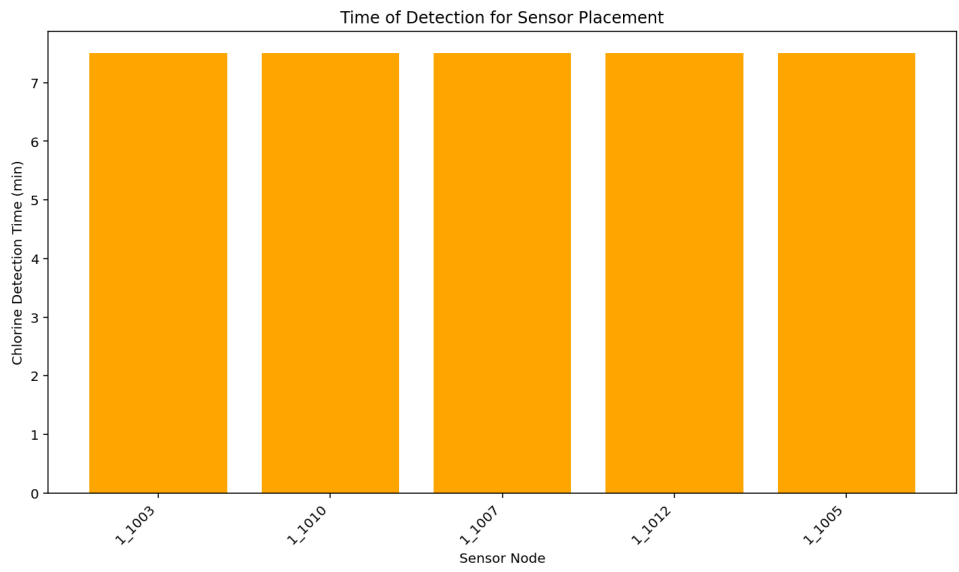

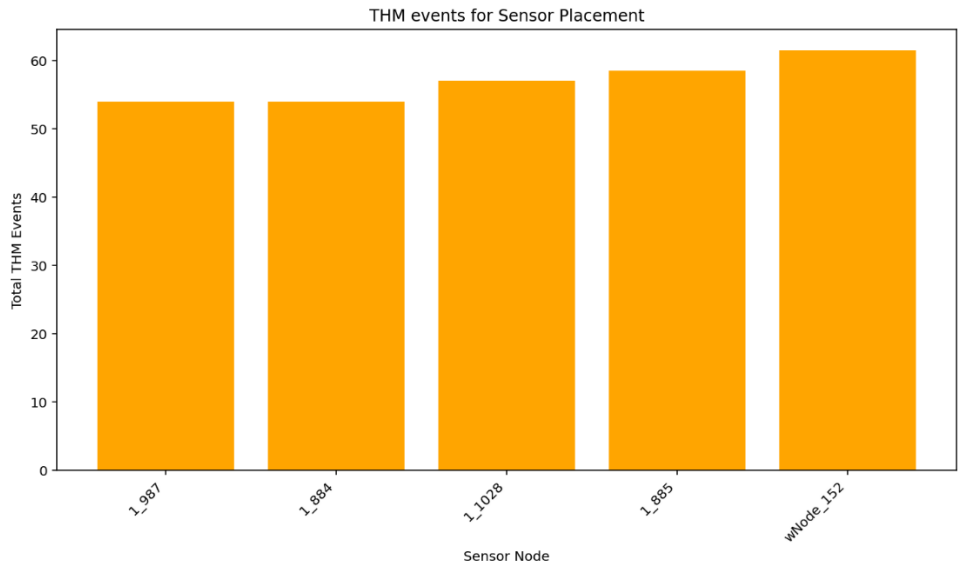
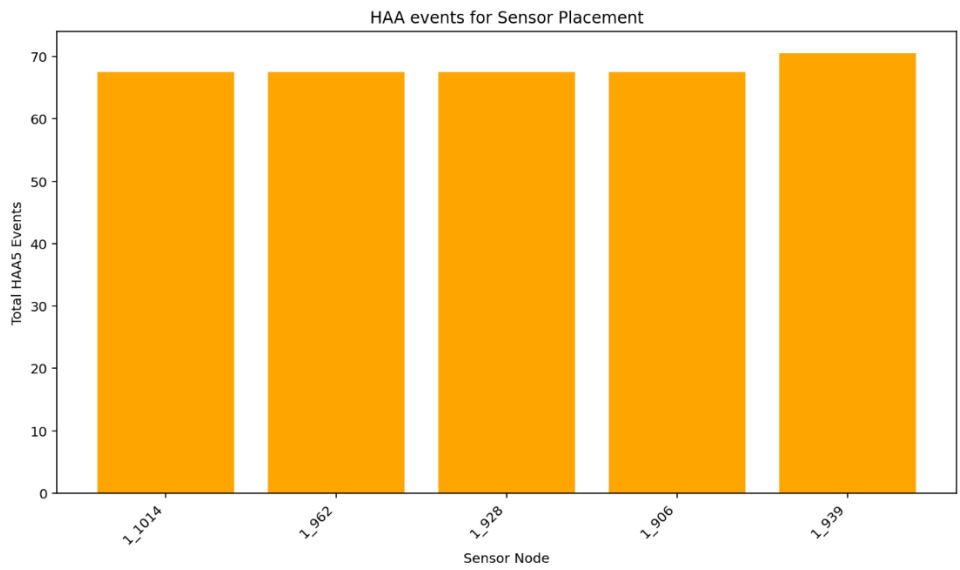

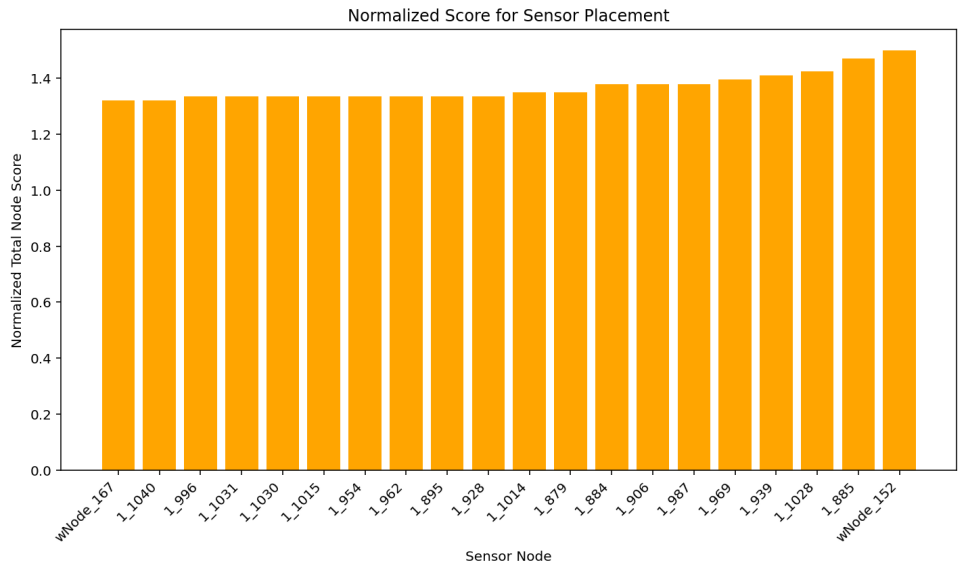
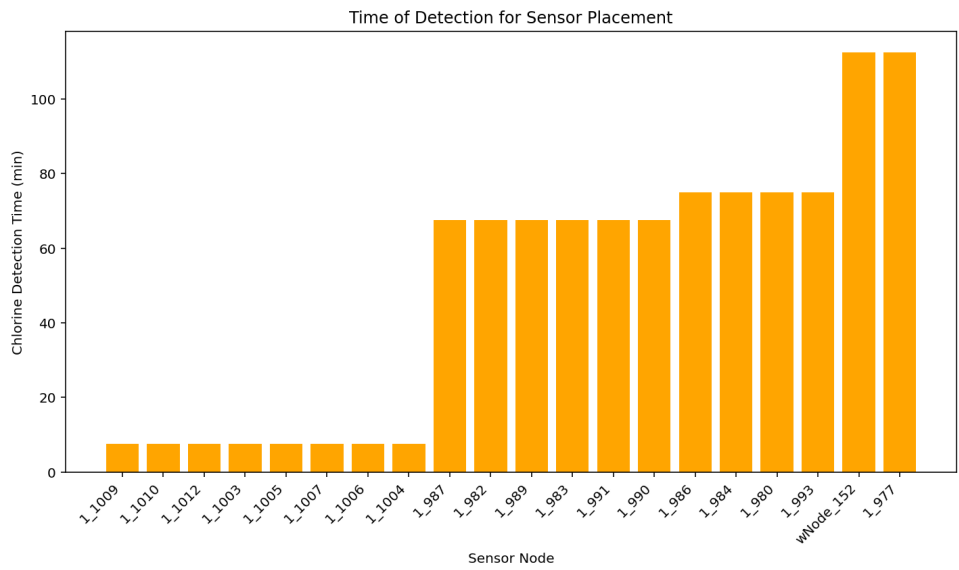

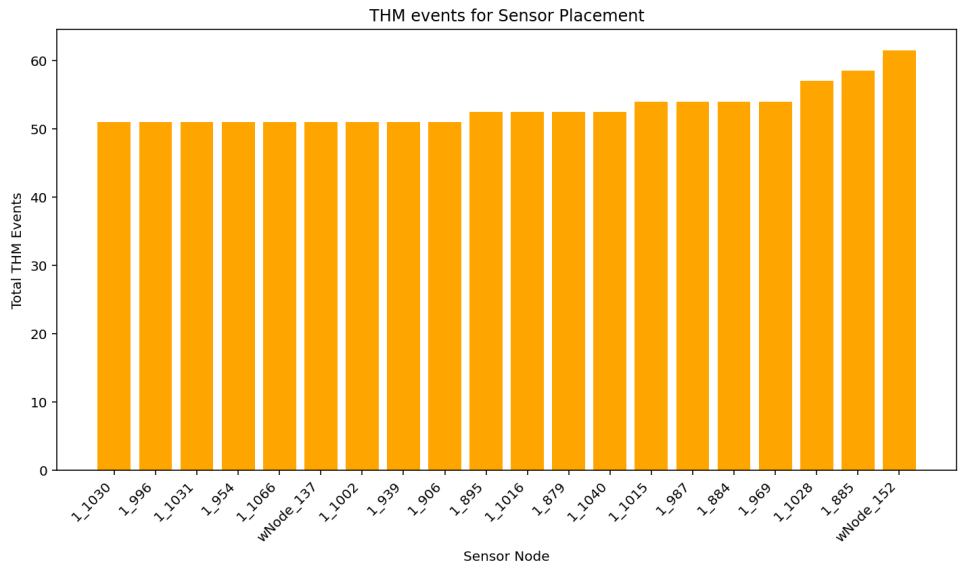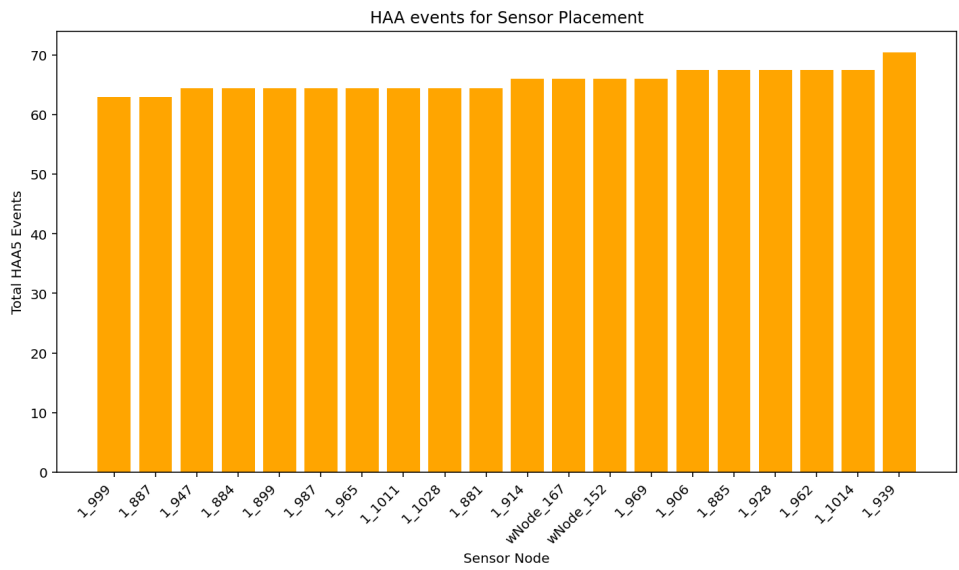